\documentclass[a4paper,fleqn]{cas-dc}
\usepackage{graphicx}
\usepackage{dcolumn}
\usepackage{bm}
\usepackage[numbers,sort&compress]{natbib}
\usepackage{etoolbox}
\apptocmd{\sloppy}{\hbadness 10000\relax}{}{}
\usepackage[utf8]{inputenc}
\usepackage{soul}

\def\tsc#1{\csdef{#1}{\textsc{\lowercase{#1}}\xspace}}
\tsc{WGM}
\tsc{QE}
\tsc{EP}
\tsc{PMS}
\tsc{BEC}
\tsc{DE}

\def\bx{\mathbf{x}}
\def\bE{\mathbf{E}}
\def\bI{\mathbf{I}}

\def\dotgamma{\dot{\gamma}}

\def\refeq#1{{{(\ref{#1})}}}

\begin{document}
\let\WriteBookmarks\relax
\def\floatpagepagefraction{1}
\def\textpagefraction{.001}

\shorttitle{Drop deformation and emulsion rheology}

\shortauthors{Reboucas et~al.}

\title [mode = title]{Modeling drop deformations and rheology of dilute to dense emulsions}                      

\author[1]{Rodrigo B. Reboucas}

\ead{rodrigor.uic.edu}

\affiliation[1]{organization={University of Illinois Chicago},
    addressline={929 W Taylor Street}, 
    city={Chicago},
    postcode={60608 IL}, 
    country={USA}}

\author[1]{Nadia N. Nikolova}

\author[1]{Vivek Sharma}
\cormark[1]
\ead{viveks@uic.edu}




\cortext[cor1]{Corresponding author}

\begin{abstract}
We highlight the current state-of-the-art in modeling emulsion rheology, ranging from dilute to jammed dense systems. We focus on analytical and numerical methods developed for calculating, computing, and tracking drop deformation en route to developing constitutive models for flowing emulsions. We identify material properties and dimensionless parameters, collate the small deformation theories and resulting expressions for viscometric quantities, list theoretical and numerical methods, and take stock of challenges for capturing connections between drop deformation, morphology, and rheology of emulsions. We highlight the substantial progress in providing quantitative descriptions of the rheological response using analytical theories, dimensional analysis, and powerful computational fluid dynamics to determine how macroscopic rheological properties emerge from microscopic features, including deformation and dynamics of non-interacting or interacting drops and molecular aspects that control the interfacial properties.
\end{abstract}




\maketitle

\section{Introduction}
\label{sec: Introduction}

Emulsions are dispersions of droplets in a continuous suspending liquid phase \cite{langevin2020emulsions, tadros2016emulsions,bibette2003emulsion,larson1999structure}. Examples of emulsions include food materials such as milk, creams, salad dressings, chocolate, and mayonnaise and cosmetics marketed as lotions and creams. Pharmaceutical formulations like certain eye drops, skin care lotions, and oral emulsions are designed such that oil phase serves as a carrier for certain hydrophobic bioactives. Dispersion of crude oil in water or vice versa during petroleum extraction or in oceans after oil spills produces petroleum emulsions. Blends of immiscible polymer solutions or melts that form dispersions of droplets in a suspending liquid are also emulsions. The immiscibility implies that the free energy of the mixing is higher than the phase separated systems formed by drops in a matrix phase \cite{larson1999structure}. The formulation of emulsions with flow properties suitable for processing, applications, and sensory perception involves quests that belong to the realm of rheology, i.e., the science of deformation and flow of simple and complex fluids (or soft matter) \cite{bibette2003emulsion,larson1999structure, macosko1994rheology,mewis2012colloidal}. Emulsion rheology is characterized by measuring the response to applied stress, strain, or strain rate, typically using specialized equipment called rheometers that are designed to create viscometric flows, or well-defined flow fields to assess flow behavior  \cite{bibette2003emulsion,larson1999structure, macosko1994rheology}. The deformability of drops, the possibility of flow within them, and their coalescence or breakup contribute to emulsion rheology \cite{langevin2020emulsions, tadros2016emulsions,bibette2003emulsion,larson1999structure} that can be quite distinct from the rheology of complex fluids containing dispersed particles, micelles, or macromolecules \cite{larson1999structure,macosko1994rheology,mewis2012colloidal,vlassopoulos2014tunable, KimMason017advances, minale2010models,barnes1994rheology,  taylor1932viscosity,oldroyd1953elastic,schowalter1968rheological,cox1969deformation,frankel1970constitutive,choi1975rheological}. Furthermore, the stability and flow behavior of emulsions depend on the composition, structure, and mechanical properties of the interface between the dispersed and continuous phases \cite{langevin2000AdvancesColloid,erni2011emulsion, erni2011deformation,fischer2007emulsion, flumerfelt1980effects}. In this contribution, we highlight how size, shape, concentration, interactions, and interfacial properties of dispersed drops influence droplet concentration-dependent variation in the rheological response of emulsions. 
 
\sloppy Processing operations such as pumping, dispensing, pouring, spreading, and even emulsion stability or shelf-life are influenced by shear viscosity $\eta$, which characterizes resistance to shear flows associated with velocity gradients perpendicular to the flow direction \cite{macosko1994rheology, larson1999structure}. Such shear flows commonly arise near solid-liquid interfaces, including pressure-driven flows through channels, and drag flows around immersed objects or near moving substrates. Most published emulsion rheology studies primarily describe the magnitude and measurement of shear viscosity, $\eta$, with a focus on stress or shear rate dependent variation \cite{larson1999structure,KimMason017advances,derkach2009rheology,barnes1994rheology,foudazi2015physical, pal2011COCIS, fischer2007emulsion}. However, stream-wise velocity gradients associated with extensional or elongational flows commonly arise in converging channels, porous media, and free surface flows, involving the formation of liquid necks that undergo capillarity-driven pinching \cite{macosko1994rheology, larson1999structure}. Studies of extensional rheology of emulsions are less common due to longstanding experimental and modeling challenges \cite{huisman2012pinch,niedzwiedz2010extensional,dinic2017pinch,nikolova2023rheology}. Extensional rheology responses profoundly influence the processing, applications, and consumer use and perception of emulsions, described in heuristic terms such as sprayability, jettability, stringiness, ropiness, and jettability \cite{huisman2012pinch,niedzwiedz2010extensional,dinic2017pinch,nikolova2023rheology}.

 Spherical drops deform into ellipsoidal shapes in response to weak velocity gradients \cite{taylor1932viscosity, oldroyd1953elastic,schowalter1968rheological, cox1969deformation,frankel1970constitutive, fischer2007emulsion} and can undergo large deformations in response to strong flows, forming slender bodies and even undergoing capillarity-driven pinching leading to breakup \cite{torza1972particle,stone1986experimental,bentley1986experimental, rallison1984deformation}. 
 Emulsification or emulsion formation, liquid blending, and emulsion rheology are three important problems that involve drop deformations in response to flow fields \cite{rallison1984deformation,fischer2007emulsion,vankova2007Part1,vankova2007Part2,ni2024AnnualReview, deroussel2001mixing}. Analytical approaches capture minor or small deformations from spherical shape, but numerical approaches are necessary to model large deformations and pinching, breakup or coalescence of drops, especially, for emulsions containing the dispersed drop phase in higher volume fractions.

 Emulsion drops deformed by velocity gradients display elasticity due to restoring stresses set by interfacial tension. After flow stops, drops can recover their unperturbed spherical shape, as it is the minimum energy configuration for a fixed drop volume \cite{oldroyd1953elastic}. The characteristic timescale for recovering this interfacial energy-favored state is called relaxation time \cite{oldroyd1953elastic} or shape or surface tension relaxation time in the emulsion rheology literature \cite{minale2010models, larson1999structure, fischer2007emulsion}. The shape relaxation time appears as viscocapillary time in interfacial fluid mechanics, including the studies of pinching, coalescence, and spreading of drops \cite{fardin2022spreading, fardin2024dynamic} as it captures the interplay of viscous and interfacial stresses. Somewhat analogous elastic response is displayed in dilute polymer solutions by polymer chains perturbed by flow, with a relaxation time as the characteristic time over which the unperturbed, entropically favored coiled state is recovered after the cessation of the flow \cite{DoiEdwards1988, larson1999structure}. In both dilute emulsions and polymer solutions, this elastic recovery of the unperturbed drop shape or coil configuration is at the heart of viscoelastic behavior, captured as modulus in stress relaxation and oscillatory shear measurements, or manifested in rod climbing or steady shear torsional rheometry as stresses due to nonzero normal stress differences \cite{larson1999structure,macosko1994rheology}.

 In non-dilute emulsions and particle suspensions, pairwise and higher order interactions and the local arrangement of discrete drops or particles constitute the microstructure that influences the flow behavior \cite{larson1999structure,macosko1994rheology, mewis2012colloidal, KimMason017advances}. In-situ visualization or monitoring of the evolution of microstructure in flow fields by optical or spectroscopic methods shows that the rearrangement of drops and the magnitude of drop deformation and orientation together determine the rheological response, including rate variation of shear viscosity and normal stress differences and amplitude and frequency dependent moduli measured using oscillatory shear \cite{fischer2007emulsion,rallison1984deformation,torza1972particle,stone1986experimental,vermant1998anisotropy}. In the jammed dense emulsion, the shear flow behavior is also influenced by deformation and flows in interconnected liquid films, leading to a yield stress that must be exceeded before flow can be observed, and typically, shear viscosity exhibits a deformation rate- or stress- dependent nonlinear response \cite{princen1989rheologyIV,mason1996yielding,larson1997elastic,cohen2014rheology}. Due to the enhanced nonlinearity and complexity of the problems, few studies explore the response of the non-dilute emulsions in extensional flows and confined flows \cite{guido2011shear}. Far fewer theoretical and simulation studies account for the influence of non-Newtonian response (rate-dependent shear and extensional viscosity, transient and nonlinear viscoelasticity) of the suspending or dispersed liquid or of the decorated, populated interface \cite{guido2011shear}.

 In this brief review, we highlight theoretical and numerical advances in modeling flows of dilute to dense jammed emulsions. The review is divided into six sections. Section \ref{sec: scope} and \ref{sec: governing equations} provide motivation, scope, brief history, definitions, transport equations, and dimensional analysis. Section \ref{sec: dilute} presents the small deformation theory and constitutive models for dilute emulsions. Section \ref{sec: concentrated} describes constitutive models and numerical methods developed for non-dilute emulsions. Section \ref{sec: dense} is a short survey of jammed dense emulsions, and Section \ref{sec: challenges} lists a few challenges and opportunities. We have included a primer on the small deformation theory in Appendices \ref{appendix: small deformation theory} and \ref{Appendix: Frankel Acrivos}, retracing steps taken by Frankel and Acrivos \cite{frankel1970constitutive} from the shape distortion tensor to the constitutive equation for dilute emulsions that can describe both transient and steady state non-Newtonian and viscoelastic response.

\section{Classifying emulsions and mapping concentration-dependent rheology}
\label{sec: scope}
\textbf{Classifying emulsions} Emulsions are classified using many criteria, ranging from the choice of dispersed and suspending liquid, interface composition, application (food,  pharmaceutical, personal care and cosmetics, petroleum)  and drop size and volume fraction range \cite{langevin2020emulsions,tadros2016emulsions, bibette2003emulsion, mcclements2017recent, zhu2020review}. Emulsions are often described on the basis of the choice of dispersed and suspending phase, oil-water or water-oil emulsions that can be obtained by mechanical mixing, phase separation, microfluidics, vapor condensation, or biologically, as in milk. Here, oil can refer to vegetable oils, crude oil (or derived oil), silicone oils, polymerizable monomers (in latex), or even organic liquids, while the water phase can be made with an aqueous solution or water-based mixed solvent. Both milk and mayo are examples of oil-water emulsions, containing water as the suspending or continuous liquid. Unlike such emulsions, water-in-water emulsions spontaneously appear as complex coacervate forms between two oppositely charged polyelectrolytes and phase separates forming emulsions that are unstable and have short shelf-life \cite{esquena2016WaterInWaterEmulsions}, though recent studies describe attempts to enhance stability against coalescence \cite{fick2023interfacial}.

Typical household emulsions, such as milk, mayonnaise, cosmetic lotions and creams, salad dressings, and fabric softeners appear milky due to scattering by drops with sizes greater than the wavelength of visible light (drop sizes, a > 1 micron). These are examples of macroemulsions and, being thermodynamically unstable, have a finite shelf-life that can be enhanced by reducing drop sizes and size dispersity, diminishing density difference, increasing the suspending fluid viscosity and manipulating drop-drop interactions \cite{langevin2020emulsions,tadros2016emulsions, bibette2003emulsion, vlassopoulos2014tunable}. Like macroemulsions, nanoemulsions (sometimes called miniemulsions) are also thermodynamically unstable, but smaller drop sizes ($a$ = 50-500 nm) and tighter size control lead to prolonged kinetic stability \cite{langevin2020emulsions,helgeson2016colloidal,gupta2017general, mcclements2012nanoemulsions}. In contrast, microemulsions that have relatively small drop sizes  ($a$ = 10-100 nm) are thermodynamically stable and appear transparent.  Classification based on interface composition: surfactant, protein, lipid, particles, polymers, or their complexes emphasizes the critical role that adsorbed species and interfacial rheology play in influencing the flow properties and stability of emulsions \cite{langevin2020emulsions,bibette2003emulsion, fischer2007emulsion}.

\begin{figure*}
	\centering
		\includegraphics[scale=.45]{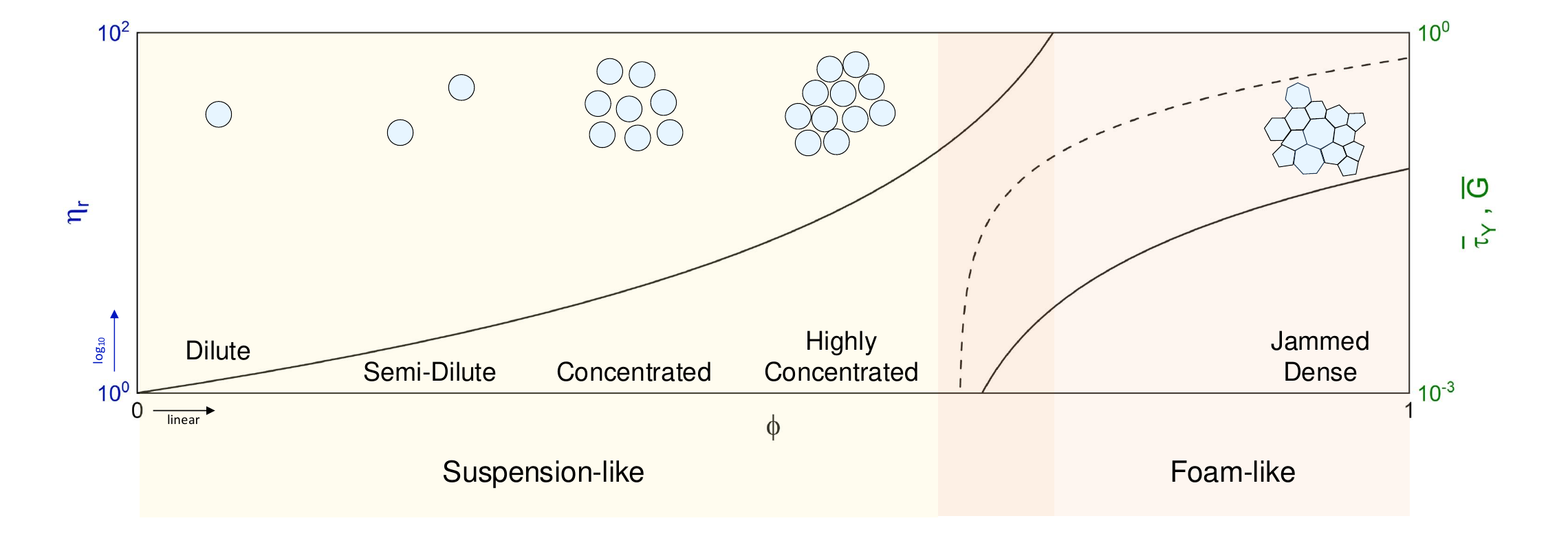}
\caption{Emulsion rheology and microstructure as a function of disperse-phase volume fraction. Representative curves show the increase in relative viscosity from dilute to highly-concentrated emulsions, and the increase in elastic modulus (dashed line) and yield stress (continuous line) for dense emulsions; $\phi$ is the disperse-phase volume fraction. Both elastic modulus and yield stress are normalized by a characteristic capillary stress $\sigma/a$.}
	\label{FIG:1}
\end{figure*}

\textbf{Concentration-dependent regimes: dilute to jammed dense} Constitutive equations that model flow properties of emulsions consider the influence of number density, interactions, and deformation of drops \cite{langevin2020emulsions,larson1999structure,KimMason017advances, barnes1994rheology, derkach2009rheology, foudazi2015physical, zhu2020review}.  The exhibited rheological behavior is considered as a linear response if the measured flow properties (stress, viscosity, or modulus) do not depend on the impelling quantities (stress, strain, or strain rate). Dilute emulsions exhibit viscosity or resistance to flow that is comparable to suspending fluid, as can be observed for animal milks, which are examples of emulsions with a relatively low $\phi$. In the dilute regime, the macroscopic properties that capture the linear viscoelastic response, including $\eta_0$ increase linearly with $\phi$. The deformation and hydrodynamics of each drop in dilute emulsion can be considered independently, by neglecting the influence of hydrodynamic and thermodynamic interactions. In semi-dilute emulsions, pairwise interactions make relative viscosity exhibit a nonlinear increase with $\phi$. In concentrated emulsions, drops are so closely packed that drop mobility and deformation become highly restricted by caging or surrounding drops. The shear viscosity exhibits a stronger non-Newtonian response for the non-dilute emulsions, and the elastic effects become progressively stronger with an increase in $\phi$. The semi-dilute to highly concentrated emulsions contain a progressively higher $\phi$ (or number density of drops), and influence of associative and repulsive inter-drop interactions and microstructure become manifest and measurable \cite{KimMason017advances, langevin2020emulsions, choi1975rheological, sanfeld2008emulsions}. 

\begin{raggedright}
Figure \ref{FIG:1} illustrates that four concentration regimes, dilute, semi-dilute, concentrated and highly concentrated emulsions, can be identified by examining the variation in relative viscosity, $\eta_{r}$ on increasing droplet volume fraction, $\phi$. Here $\eta_{r}=\eta/\mu$ representing the emulsion's zero shear viscosity scaled with suspending fluid viscosity, $\mu$. Viscosity increases with $\phi$ substantially in the highly concentrated regime, qualitatively emulating the behavior of rigid particle suspensions, where viscosity diverges close to maximum volume fraction \cite{morris2020annualRev,abhi2020PRL, vlassopoulos2014tunable}. Due to deformability of drops, droplet volume fraction can be increased further leading to the jammed dense emulsion regime. As volume fraction of drops lies beyond the maximum packing fraction for spherical or ellipsoidal particles, jammed dense emulsions contain polygonal-shaped drops separated by interconnected liquid films with a foam-like microstructure. Mayonnaise, an egg-based emulsion of vegetable oil droplets suspended in a aqueous medium \cite{nikolova2023rheology}, is an example of jammed dense emulsion containing closely-packed, polygonal drops, with volume fraction of the drop phase between $\sim65\%-80\%$. Such dense emulsions display yield stress, $\tau_Y$,  and elastic modulus, $G$, that increases with volume fraction \cite{langevin2020emulsions, larson1999structure,KimMason017advances,foudazi2015physical}.The variation in yield stress and modulus scaled by capillary pressure  is illustrated in the Figure \ref{FIG:1} for jammed dense emulsions. Though it is well-established that increasing drop volume fraction leads to a transition from suspension-like to foam-like behavior as shown schematically, for emulsion drops and for deformable particles, the transition region depends on many factors including size and shape, size dispersity, interactions, and mechanisms underlying the deformability of the dispersed phase \cite{KimMason017advances, datta2011rheology, cao2021rheology, langevin2020emulsions, vlassopoulos2014tunable}.
\end{raggedright}

\textbf{Highlights from ninety years of analytical models for emulsion rheology}  The review encompasses and primarily highlights models that rely on small deformation theory, a perturbation calculation for weak deviations about a spherical shape that are apt for dilute emulsions, and provide insights into rheology of nondilute emulsions \cite{taylor1932viscosity, chaffey1967second, cox1969deformation, frankel1970constitutive, ramachandran2012effect, narsimhan2019shape, minale2010models, oldroyd1953elastic, barnes1994rheology}. In 1906, Einstein connected the microhydrodynamics calculation of the flow around dispersed particles to the estimation of viscosity of a dilute suspension of hard spheres \cite{einstein1906, einstein1911}. Subsequently, Taylor (1932) first analyzed drop deformation in the presence of flow \cite{taylor1932viscosity} and generalized Einstein's theory \cite{einstein1906, einstein1911} to describe the viscosity of dilute emulsions by accounting for internal circulation. Decades later, Schowalter, Chaffey and Brenner (1968) \cite{schowalter1968rheological} extended the model to suggest the existence of normal stress components, but their model reveals no viscosity variation due to drop deformation. Frankel and Acrivos (1970) \cite{frankel1970constitutive}, and Barthès-Biesel and Acrivos (1972) \cite{barthes1973rheology} developed constitutive equations for dilute emulsions that describe the response to transient flows. Choi and Schowalter (1975) \cite{choi1975rheological} carried out the extension to semi-dilute solutions, whereas Princen and Kiss (1980s) \cite{princen1989rheologyIV} showed the connection between yield stress or elastic modulus and surface tension for dense emulsions and foams. Flumerfelt (1980) first examined the influence of interfacial tension variation as well as dilatational and shear interfacial viscosity on drop deformation in the small deformation limit \cite{flumerfelt1980effects}. Later, Leal, Stone and coworkers performed more extensive examination in the limit of large deformation, including the influence of surfactants \cite{stone1990effects, ramachandran2012effect, bentley1986experimental, fischer2007emulsion, erni2011emulsion}.

Barthès-Biesel (1980) began the examination of deformation and rheology of capsules, defined as viscous drops covered with elastic membranes, and showed that the combination of liquid-like interior enclosed within a solid-like shell leads to behaviors that cannot be inferred from suspension of hard spheres or emulsion containing drops with Newtonian interfaces \cite{barthes1980motion, barthes1981time, barthes2011modeling, barthes2016motion}. Oldroyd (1954, 1955) \cite{oldroyd1953elastic, oldroyd1955effect} presented the first attempt to describe the rheology of nondilute emulsions by adopting the effective medium theory proposed in 1946 by Fröhlich and Sack for the dispersion of deformable particles \cite{frohlich1946theory}. Oldroyd also introduced a tensorial framework to capture the complex viscoelastic response of emulsions with appropriate attention to frame invariance. Starting with Taylor's discussion of drop deformation \cite{taylor1932viscosity} or with Oldroyd’s framework \cite{oldroyd1953elastic, oldroyd1955effect}, a large number of analytical and continuum models have emerged, which incorporate the interplay of drop deformation, interactions, breakup and coalescence processes and rely on numerical and computation approaches, especially for connecting the microstructure and rheology of nondilute and dense emulsions. We provide a selective (and incomplete) but pragmatic overview of the theoretical framework necessary for modeling emulsion rheology.

\textbf{Scope of this review} We provide a brief synopsis of the small-deformation theories based on perturbation methods that are used to capture drop deformation and the rheological response of dilute emulsions to viscometric flows. As dilute emulsions contain non-interacting drops, their shear rheology response under weak flows, including shear viscosity, can be computed by recognizing that contributions from each drop, the mildly perturbed drop by the imposed flow, must be added to those by the suspending fluid \cite{mwasame2017macroscopic,schowalter1968rheological,frankel1970constitutive,choi1975rheological,pal1992rheology,kennedy1994motion,li1997effect,guido2004dynamics,aggarwal2008rheology,vlahovska2009small,minale2010models,ramachandran2012effect,oliveira2015emulsion,narsimhan2019shape, cox1969deformation}. Deviations from small-deformation conditions are captured by numerical simulations. We organized the discussion according to the composition of the droplet interface in clean drops and surfactant-covered drops modeled as droplets with surface viscosity. Quantitative descriptions of the rheological response for non-dilute emulsions rely on supplementing analytical theories with computational fluid dynamics to determine the contributions from deformation and dynamics of non-interacting or interacting drops and molecular aspects that control the interfacial properties. We tabulate different methods and highlight their key findings. As the macroscopic rheological response of emulsions is often compared with the expectations based on constitutive models developed for suspensions of undeformable particles, we include suitable references for completeness \cite{einstein1906,einstein1911, larson1999structure, morris2020annualRev, mewis2012colloidal,vlassopoulos2014tunable, macosko1994rheology}. 

In this opinion, we exclude discussions relevant to emulsification and highly nonlinear flows of emulsions \cite{vankova2007Part1,vankova2007Part2,ni2024AnnualReview}. We briefly mention the influence of viscoelastic interfaces, drops, or suspending liquids on the emulsion rheology and for the sake of brevity, highlight articles and reviews that detail recent progress and open questions \cite{escott2024rheology,villone2019dynamics, guido2011shear, minale2010models, fischer2007emulsion, langevin2020emulsions}. Likewise, we exclude studies on capsule suspensions \cite{barthes2011modeling, barthes2016motion}, wall effects, and the influence of external force fields on droplet topology and emulsion microstructure \cite{miksis1981shape,cristini2004review,bazhlekov2006numerical,vananroye2008microconfined,janssen2010generalized,guido2011shear,cunha2018field,roure2023numerical}. We cite a paucity of datasets and the immensity of challenges involved in theoretical and experimental studies of the extensional rheology response as a reason for excluding a detailed exposition of the few published studies, including our own \cite{nikolova2023rheology, niedzwiedz2010extensional, rallison1980note, kibbelaar2023towards}. We do not cover studies on Pickering emulsions, water-in-water emulsions, microemulsions, and nanoemulsions, and recommend some recent reviews \cite{langevin2020emulsions, helgeson2016colloidal, gupta2017general, hermes2013yielding,fick2023interfacial, dardelle2014three}. We exclude any discussion of rheometry techniques and measured rheological response of emulsions or interfaces enriched with adsorbed species, but we anticipate that the references included here can be used as a guide for the road not taken \cite{langevin2020emulsions, tadros2016emulsions, larson1999structure,erni2007viscoelastic, KimMason017advances, derkach2009rheology, macosko1994rheology, erni2005deformation, erni2011deformation, sharma2011rheology, wasan1991book, bibette2003emulsion, fischer2007emulsion, flumerfelt1980effects, foudazi2015physical}. Although capillary pressure, interfacial rheology, disjoining pressure (contributed by intermolecular and surface forces), and bulk rheology of two liquids all influence drainage flows in thin liquid films that separate any droplet pairs and therefore influence emulsion stability and rheology, a comprehensive description of these remains an open challenge \cite{langevin2020emulsions, KimMason017advances, sanfeld2008emulsions, chatzigiannakis2021thin}. However, we plan to highlight reviews, monographs, articles, and textbooks that form essential reading for appreciating the state-of-the-art understanding and progress in experimental, theoretical, and computational studies of emulsion rheology \cite{larson1999structure,KimMason017advances,derkach2009rheology,foudazi2015physical,barnes1994rheology, minale2010models, fischer2007emulsion, bibette2003emulsion,  flumerfelt1980effects, erni2011emulsion}.

\section{Emulsion microhydrodynamics: the governing equations and scaling} \label{sec: governing equations}

\subsection{Governing equations and boundary conditions}
Emulsions are structured two-phase fluids composed of droplets of density $\rho +\Delta \rho$ and viscosity $\lambda \mu$  suspended in a continuous-phase fluid of density $\rho$ and viscosity $\mu$. If both the dispersed and the continuous phases are Newtonian, incompressible fluids, and interface is also Newtonian and slip or dissipation free, the only additional material parameter needed is the interfacial tension that depends on the two liquids chosen. In the continuum limit, and in the absence of body-force torques, the linear momentum and mass conservation equations are

\begin{equation}
   Re \left(\frac{\partial \mathbf{u}}{\partial t} + \mathbf{u} \cdot \nabla \mathbf{u}\right) = \nabla \cdot \mathbf{\Sigma} \,; \quad \nabla \cdot \mathbf{u}=0 \,,
\end{equation}
where $\mathbf{u}$ is the velocity field averaged over a continuum volume of fluid, $\mathbf{\Sigma}$ is volume-averaged stress tensor in the emulsion, and $Re={u}L/\nu$ is the macroscopic Reynolds number defined as ratio of convective to diffusive velocity (or ratio of inertial to viscous stress) \cite{fardin2024dynamic}. Here we are assuming that variations of an emulsion macroscopic flow occur over a characteristic length scale $L$, such that $a/L \ll 1$, where $a$ is the average, undisturbed droplet size. Even for emulsions with Newtonian flow behavior for both phases, the  emulsion exhibit non-Newtonian rheological behavior (e.g., shear thinning and normal stress differences) due to the interplay of droplet-level deformation and relaxation, interfacial dynamics, and interdrop interactions leading to an anisotropic emulsion microstruture in response to imposed bulk stresses \cite{loewenberg1998conc}.

In most applications where emulsions play a key role, droplet size is within the few nanometer to few  micron scale such that the local Reynolds number defined in terms of the local shear rate and particle size is $Re_{local}=Re (a/L)^2$ provided that $a/L \ll 1$. Hence, the microhydrodynamics at the droplet level are governed by the low-Reynolds-number flow equations, 
\begin{equation}
   \label{eq Stokes continuous}
    \mu \nabla^2\mathbf{u} -\nabla p +\rho \mathbf{g} = 0 ; \quad \nabla \cdot \mathbf{u}=0 
\end{equation}
\begin{equation}
    \label{eq Stokes drop}
    \lambda \mu \nabla^2\mathbf{u'} -\nabla p' +(\rho +\Delta \rho) \mathbf{g} = 0 ; \quad \nabla \cdot \mathbf{u'}=0 
\end{equation}
where the primes denote quantities associated with the drop phase, $\mathbf{g}$ is the gravitational acceleration, and $p$ is the mechanical pressure. Equations \refeq{eq Stokes continuous}-\refeq{eq Stokes drop} are valid everywhere expect at the droplet interface denoted by $S$. Often models assume that the suspending liquid is density matched with the droplet or dispersed phase.  

Boundary conditions encompass, typically, an imposed flow field
\begin{equation}
    \label{BC far field}
    \mathbf{u} \to \mathbf{u}^{\infty}  \quad \text{as} \quad \vert \mathbf{x}\vert \to \infty \,,
\end{equation}
where $\mathbf{x}$ is the position vector measured from the droplet center. In a general case where the droplet interface is covered with a slip layer of a macromolecule, the Navier-slip condition is used
\begin{equation}
    \label{eq BC velocity jump}
    \mathbf{u}-\mathbf{u'}=\alpha(\mathbf{I}-\mathbf{n}\mathbf{n})\cdot (\mathbf{T}\cdot \mathbf{n}) \quad \text{for} \quad \mathbf{x}_S \in S\,,
\end{equation}
where $\mathbf{T}$ is the local Newtonian stress tensor, $(\mathbf{I}-\mathbf{n}\mathbf{n})\cdot (\mathbf{T}\cdot \mathbf{n})$ is the tangential component of the stress vector $\mathbf{T}\cdot \mathbf{n}$ at the interface, $\mathbf{x}_S$ is a point at the droplet surface, and $\alpha$ is a slip coefficient. Generally, the velocity at the interface is continuous and $\alpha=0$. The traction jump at the interface is given by 
\begin{equation}
    \label{eq BC traction jump}
    [\mathbf{n}\cdot \mathbf{T}]_S=\left(2 H \sigma + \Delta \rho \mathbf{g} \cdot \mathbf{x}\right)\mathbf{n} - \nabla_S \sigma  \quad \text{for}\quad \mathbf{x}_S \in S\,,  
\end{equation}
where $[.]_S$ denotes a jump of the bracketed quantity across the interface, $\nabla_S = (\mathbf{I} - \mathbf{n}\mathbf{n}) \cdot \nabla$ is the surface gradient operator, and $\sigma$ is the interfacial tension coefficient which may vary along the droplet interface in response to gradients in temperature or in-homogeneous distribution of surfactant molecules. The mean curvature $H$ is computed using
\begin{equation}
    \label{eq mean curvature}
    H=\frac{1}{2}\nabla_S \cdot \mathbf{n}
\end{equation}
In such cases, an equation of state and an evolution equation for surfactant concentration, $\Gamma$, are needed for closure \cite{stone1990simple,eggleton1998adsorption}. Several adsorption isotherms that outline how surface tension varies with change in interfacial concentration of surfactants can be used. We direct the interested reader to Table 1 of Ref.~\cite{manikantan2020surfactant} for a comprehensive list. Here, we illustrate methodology using  the non-linear Langmuir equation of state,
\begin{equation}
    \label{eq Langmuir EoS}
    \sigma(\Gamma)= \sigma_0 +R T \Gamma_{\infty} \ln \left(1-\frac{\Gamma}{\Gamma_{\infty}}\right) \,,
\end{equation}
where $\sigma_0$ is the surface tension of the clean (surfactant-free) interface, $R$ is the ideal gas constant, $T$ is the absolute temperature, and $\Gamma_{\infty}$ is the maximum packing concentration of surfactant molecules in a monolayer. In the absence of flow and after surfactant adsorption occurs for a sufficient time, there is an equilibrium surface tension $\sigma_{eq}$ at which the equilibrium surface pressure $\Pi_{eq} = \sigma_0 -\sigma_{eq}$ is defined for a given equilibrium surfactant concentration, $\Gamma_{eq}$ \cite{eggleton1999insoluble}. The ratio $\Gamma_{eq}/\Gamma_{\infty}$ known as surface coverage indicates the initial fraction of the interface covered with surfactants. 

In the limit of dilute bulk concentration of surfactants, the adsorption kinetics and bulk surfactant diffusion are slow compared to local-convective-flow time scales, such that the surfactant layer at the interface is approximately insoluble and follows a time-dependent convection-diffusion equation \cite{leal2007advanced},
\begin{equation}
    \label{eq surfactant evolution}
    \frac{\partial \Gamma}{\partial t} + \nabla_S \cdot (\Gamma \mathbf{u}_S) -D_S \nabla^2_S \Gamma +2 H \Gamma(\mathbf{u}\cdot \mathbf{n})=0 \,,
\end{equation}
where $\mathbf{u}_S=(\mathbf{I}-\mathbf{n}\mathbf{n})\cdot \mathbf{u}$ is the tangential component of velocity at the interface, and $D_S$ is the surfactant interfacial diffusivity. The second term in Eq.~\refeq{eq surfactant evolution} represents surface convection, the third indicates surface diffusion, and the last represents surface dilution due to local changes in interfacial area or surface dilatation. 

The evolution of the droplet interface is captured by the kinematic boundary condition,
\begin{equation}
    \label{eq BC kinematic} 
    \frac{d \mathbf{x_s}}{dt}=\mathbf{n}(\mathbf{u} \cdot \mathbf{n}) \,.
\end{equation}
 
\subsection{Relevant physicochemical parameters, scales, and dimensionless groups} \label{sec: relevant parameters}
A characteristic length scale for describing deformation, breakup or coalescence of drops, is the undeformed drop size, $a$. A possible characteristic time scale can be defined in terms of the capillary relaxation time or shape relaxation time, written as:
\begin{equation}
    \label{eq relaxation time}
    \tau_{\sigma} = \mu a/\sigma_{eq}, \; or \; \tau_{\sigma} =\lambda\mu a/\sigma_{eq} ,
\end{equation}
as the larger of the two viscosities determines the time period for shape relaxation, the characteristic time scale for $\lambda \gg 1$ is defined as $\tau_{\sigma} = \lambda\mu a/\sigma_{eq}$ \cite{frankel1970constitutive, rallison1984deformation}. Otherwise $\tau_{\sigma} = \mu a/\sigma_{eq}$ is typically used, and $\tau_{\sigma}$, also called viscocapillary time, captures the time required to traverse a distance comparable to drop size, with an intrinsic capillary velocity, $\sigma_{eq}/\mu$ set by the ratio of two physicochemical parameters or material properties: interfacial tension and viscosity \cite{fardin2024dynamic}. The two material parameters can be used to estimate characteristic scale for pressures or stresses, as follows. The ratio $\sigma_{eq}/a$, provides an estimate for capillary stress, whereas $\mu \dot{\gamma}$ estimates the viscous stress.

Assuming a neutrally-buoyant drop ($\Delta \rho=0$) in an imposed linear flow field where $\mathbf{u}^{\infty} \sim \mathbf{x} \cdot \nabla \mathbf{u}$, the characteristic time scale for the flow is $\tau_f =\dot{\gamma}^{-1}$, where $\dot{\gamma}$ is the magnitude of the local velocity gradient. A typical timescale for droplet deformation in shear is $\tau_d \sim \tau_f = \dot{\gamma}^{-1}$. Setting the undeformed drop size, $a$, as the characteristic length scale, a natural choice for the characteristic velocity is $\dot{\gamma} a$ and hence, from Eqs. \refeq{eq Stokes continuous}-\refeq{eq Stokes drop} the pressures inside and outside of the droplet scale as $\mu \dot{\gamma}$ and $\lambda \mu \dot{\gamma}$, respectively. The choices of characteristic time, length and stress/pressure scales determine the form of dimensionless equations and boundary conditions obtained after a nondimensionalization of Eqs.~\refeq{eq Stokes continuous}-\refeq{eq BC kinematic}.

The dimensionless ratio of viscous and capillary stresses, is defined as the capillary number
\begin{equation}
    \label{eq Ca number}
    Ca=\frac{\mu \dot{\gamma}}{\sigma_{eq}/a}=\frac{\dot{\gamma}a}{\sigma_{eq}/\mu}=\frac{\tau_{\sigma}}{\tau_d}\,.
\end{equation}
Alternatively, $Ca$ equals the ratio of imposed flow velocity, $\dot{\gamma} a$ to intrinsic capillary velocity, $\sigma_{eq}/\mu$.  The capillary number can be written equivalently as the ratio of capillary relaxation time to deformation time. Since $Ca$ is also a product of relaxation time, $\tau_{\sigma}$ and deformation rate ($\dot{\gamma}$ for shear), it captures the flow strength in a fashion reminiscent of Weissenberg number $Wi= \dot{\gamma}\tau_{1}$ used in polymer rheology, with $\tau_{1}$ representing the longest relaxation time. Thus, $Ca$ captures the relative magnitude of stress, velocity, and flow strength for calibrating the influence of applied flow conditions on drop deformation and dynamics. Again, for $\lambda \gg 1$, the $Ca$ values should be computed by considering $\tau_{\sigma} = \lambda\mu a/\sigma_{eq}$ as the shape relaxation time \cite{frankel1970constitutive, rallison1984deformation}.

Two additional dimensionless groups are written as the ratio of stresses or pressures. The Bond number, $Bo$ captures the ratio of hydrostatic to capillary pressures, relevant to determining buoyancy-driven motion and the influence of gravity on the shape and deformation of drops. The Marangoni number, $Ma$, is a ratio between restoring Marangoni stresses $\Delta \sigma/a$ that arise due to surface tension variation, $\Delta \sigma$ and distorting viscous stresses,
\begin{equation}
    \label{eq Bo and Ma numbers}
    Bo=\frac{\Delta \rho g a}{\sigma_{eq}/a}\,,  \quad Ma^{-1}= \frac{\mu \dot{\gamma}}{\Delta \sigma/a}\,.
\end{equation}

 If the origin of the Marangoni stress is a non-uniform surfactant contribution, then the characteristic magnitude of surface-tension variation equals the magnitude of surface compression modulus $\Delta \sigma=-\Gamma_{eq} \left(\partial \sigma/\partial \Gamma \right)_{\Gamma=\Gamma_{eq}}$  that arises from perturbations about the equilibrium surface concentration, $\Gamma_{eq}$. The dimensionless ratio of $\Delta \sigma$ to ${\sigma}_{eq}$ represented by $\beta$ is a surface elasticity parameter \cite{li1997effect,vlahovska2009small},
 \begin{equation}
    \label{eq E and Pe numbers}
     \quad \beta=\frac{\Delta \sigma}{\sigma_{eq}}=CaMa\,,\quad Pe_{S}=\frac{\dot{\gamma}a^2}{D_S}\,,
\end{equation}
where, $Pe_S$ is the surface Péclet number denoting the relative balance between surfactant convection and diffusion along the interface. Modeling emulsification by mechanical methods \cite{vankova2007Part1, vankova2007Part2, haakansson2019emulsion, haakansson2023emulsifier} can sometimes require the evaluation of inertial effects using the characteristic inertial pressure estimated as ${\rho}U^2$. For example, Reynolds number, $Re=\rho U^2 /(\mu U/a)$ and Weber number, $We=\rho U^2/(\sigma/a) $ are defined as the ratio of inertial pressure to viscous and capillary stress, respectively \cite{fardin2024dynamic}.

Dissipative effects due shear and dilatational surface viscosity may affect the dynamics of droplets in flows \cite{fischer2007emulsion}. Two dimensionless Boussinesq numbers that capture the interplay between bulk viscous stresses and dissipative interfacial stresses, are defined as
\begin{equation}
    \label{eq Boussinesq numbers}
    Bq_{s}=\frac{\mu_s}{\mu a} \,, \quad Bq_{d}=\frac{\mu_d}{\mu a}
\end{equation}
for shear surface viscosity, $\mu_s$, and dilatational viscosity, $\mu_d$, respectively. In such cases, the right-hand side of the traction jump boundary condition in Eq.~\refeq{eq BC traction jump} is augmented by an additive interfacial-viscous traction of form, $\nabla_S \cdot \tau_S$, obeying the deviatoric part of the Boussinesq-Scriven constitutive law for Newtonian interfaces \cite{wasan1991book,schwalbe2011interfacial,jaensson2021computational, erni2011deformation},
\begin{equation}
    \label{eq Boussinesq-Scriven}
   \mathbf{\tau}^s=2 \mu_s \mathbf{E}_s +(\mu_d-\mu_s)(\mathbf{I}_S : \mathbf{E}_S)\mathbf{I}_S\,,
\end{equation}
where $\mathbf{E}_S = \frac{1}{2} \left[\nabla_S \mathbf{u} \cdot \mathbf{I}_S+\mathbf{I}_S \cdot (\nabla_S \mathbf{u})^T\right]$ is the surface rate of deformation tensor, and $\mathbf{I}_S=\mathbf{I}-\mathbf{n} \mathbf{n}$ is a surface projector tensor. Consistent with the traction jump in Eq.~\refeq{eq BC traction jump}, normalizing Eq.~\refeq{eq Boussinesq-Scriven} by a characteristic surface stress $\mu \dot{\gamma} a$, characteristic length $a$, and velocity $\dot{\gamma}a$ yields the dimensionless Boussinesq numbers in Eq.~\refeq{eq Boussinesq numbers}. 

Incorporating surface viscosity can alter the interfacial force balance in Eq. \refeq{eq BC traction jump} and interfacial transport of surface-active entities at complex interfaces. Gradients in surface tension ($\nabla_S \sigma$) generate Marangoni stresses. Interfacial shear viscosity characterizes the resistance to interfacial shear flow, and surface dilatational viscosity captures the resistance to dilatational effects that can influence coalescence \cite{pozrikidis1994effects,luo2019influence,narsimhan2019shape,herrada2022stability}. Interfacial concentration and interaction between adsorbed molecules (and macromolecules) influence interfacial tension, $\sigma$ and surface pressure, $\Pi=\sigma_0-\sigma$ that depends on the surface tension reduction compared to value at a clean interface, $\sigma_0$.

Several studies suggest that surface viscosity depends exponentially on surface pressure \cite{kim2011interfacial,kim2013effect,samaniuk2014micro,hermans2014interfacial,manikantan2020surfactant}
\begin{equation}
    \label{eq viscous pressure}
    \mu_{i}=\mu_{i,eq}\exp\left(\frac{\Pi-\Pi_{eq}}{\Pi_c}\right)\,,
\end{equation}
where $i=s,d$ identify shear and dilatational viscosities, $\mu_{i,eq}$ and $\Pi_{eq}$ are the equilibrium surface viscosity and surface pressure, respectively, and $\Pi_c$ is a characteristic scale of surface pressure variations. Positive values of $\Pi_c$ indicate $\Pi$-thickening surfactants, while negative values are used for $\Pi$-thinning surfactants. The relation between surfactant transport and surface viscous stresses is given by combining Eqs.~\refeq{eq Langmuir EoS} and \refeq{eq viscous pressure} yielding the surfactant-concentration-dependent Boussinesq numbers, defined as follows

\begin{equation}
    \label{eq Boussinesq pressure}
    Bq_{i}=Bq_{i,eq}\left(\frac{1-\hat{\Gamma}_{eq}}{1-\hat{\Gamma}}\right)^{\beta/\hat{\Pi}_c}\,,
\end{equation}
where $i=s,d$ indicate the type of surface viscosity, $Bq_{i,eq}$ is a reference equilibrium value, $\hat{\Pi}_c=\Pi_c/\sigma_{eq}$, $\hat{\Gamma}=\Gamma/\Gamma_{\infty}$, $\hat{\Gamma}_{eq}=\Gamma_{eq}/\Gamma_{\infty}$, and $\beta$ is the elasticity parameter. Typically, the ratio of dilatational to surface viscosity $\lambda_{ds}$ is used to study the relative importance of both surface viscosities. 

\sloppy
Emulsions of droplets with slip-boundaries have been used to model the rheology of emulsions of immiscible polymer blends, where the slip coefficient is defined by the ratio of the interfacial thickness and some isotropic interfacial viscosity \cite{ramachandran2012effect,salac2023ModelingSlipDrops, cardinaels2016morphology}. Non-dimensionalizing Eq.~\refeq{eq BC velocity jump} yields a dimensionless slip coefficient $\bar \alpha=\alpha/(\mu a)$.

Emulsions with one or both phases as non-Newtonian require additional parameters and considerations, which depend on the choice of constitutive model made for capturing one or more features typical of non-Newtonian behavior: rate-dependent shear and extensional viscosity, first and second normal stress difference, and relaxation time. Even in the simplest case of the second-order fluid model for both phases, two normal stress differences ${N_{1i}}$ and ${N_{2i}}$ each arise for dispersed ($i=d$) and suspending ($i=s$) phase, creating at least four additional dimensionless parameters:
\begin{equation}
    \label{eq VE dimensionless}
    {N_{1i}}a/\sigma_{eq}, \: N_{2s}/N_{1s}, \: N_{2d}/N_{1d}, \: De_{i}=\tau_{i}/\tau_{\sigma},
\end{equation}
where $\tau_{i}={N_{1i}/\Sigma_{12i}}$ can be used for defining the relaxation time for suspending or disperse phase, in which case, if we define $\bar N_{1i}={N_{1i}}a/\sigma_{eq}$ then $De_{i}=\bar N_{1i}/{Ca}^2$, \cite{greco2002drop, minale2010models, guido2011shear}. The relaxation time for viscoelastic fluid phase can be alternatively determined using the linear viscoelastic response measured in oscillatory shear, stress relaxation, dynamic light scattering or capillarity-based extensional rheology, and each response captures aspects of non-Newtonian response that need not correlate directly with the first normal stress difference.

\subsection{Emulsion macroscopic stress}
The continuum, macroscopic volume-averaged stress in Eq.~\refeq{eq Stokes continuous} for an emulsion where both dispersed and suspending fluids are Newtonian is 
\begin{equation}
    \label{eq average stress suspension}
    \mathbf{\Sigma}=\mathbf{\Sigma}^0 + \phi \mathbf{\Sigma}^p \,,
\end{equation}
where $\phi$ is the drop-phase volume fraction, and $\langle . \rangle$ denote the volume-average of the quantity in brackets. Here $\mathbf{\Sigma}^0=-\langle p\rangle \bI +2 \mu \langle \bE \rangle$ is the Newtonian stress contribution from the continuous phase. In analogy with a particulate system \cite{batchelor1967introduction}, the extra stress in an emulsion due to the dispersed droplets can be determined using the following expression:
\begin{equation}
    \label{eq particle stress}
    \mathbf{\Sigma}^p=\frac{3}{4 \pi a^3} \frac{1}{N} \sum_{\alpha=1}^{N} \mathbf{S}^{\alpha}\,,
\end{equation}
where the sum accounts for the stress contribution of each one of the $N$ drops in emulsion (or particles in suspension) given by the Landau-Batchelor tensor \cite{batchelor1970stress} defined as
\begin{equation}
    \label{eq stress Landau-Batchelor}
    \mathbf{S}^{\alpha}_{ij}= \int_{S} \left[ 
 \left(\Delta \mathbf{f}\right)_i x_j +\mu (\lambda-1) (u_i n_j + n_i u_j) \right] dS
 ,.
\end{equation}
The Landau-Batchelor tensor depends on the surface traction and the velocity distribution over the particle surface, where local low-Reynolds number conditions hold and no external torques are applied. In the limit of a sharp fluid interface, $\mathbf{\Sigma}\cdot \mathbf{n}\to \Delta \mathbf{f}$ captures the stress jump across the interface defined in Eq.~\refeq{eq BC traction jump}. For example, considering clean, neutrally buoyant droplets, $\Delta \mathbf{f}=2 H \sigma \mathbf{n}$. 

Thus, the connection between microscopic behavior and macroscopic rheology is embedded in the definition of the macroscopic particle-stress contribution  $\mathbf{\Sigma}^p$ given by equation Eq.~\refeq{eq particle stress}. The stress jump across the interface, $\Delta \mathbf{f}$, captures the microscale physics while $\mathbf{\Sigma}^p$ accounts for the contribution of the dispersed phase. The emulsion shear rheology is defined by a shear viscosity $\Sigma_{12}$ and first- and second-normal stress differences that arise from contributions of the dispersed phase only, \cite{bird1987dynamics}
\begin{equation}
    \label{eq N1 emulsion}
    N_1=\phi N^p_1=\phi(\Sigma^p_{11}-\Sigma^p_{22})\,,
\end{equation}
\begin{equation}
    \label{eq N2 emulsion}
    N_2=\phi N^p_2=\phi(\Sigma^p_{22}-\Sigma^p_{33})\,.
\end{equation}
Relative viscosity, $\eta_{r}\equiv (\eta/\mu)$ that equals emulsion viscosity $\eta$ scaled by the suspending fluid viscosity $\mu$ can be defined in terms of $\bar \Sigma_{12}$the dimensionless form of extra stress due to added particles or drops as given by Eq.~\refeq{eq particle stress}. Thus, the relative viscosity has the following form
\begin{equation}
    \label{eq effective viscosity}
    \eta_{r}=1+\phi Ca^{-1} \bar{\mathbf{\Sigma}}^p_{12}\,,
\end{equation}
where
\begin{equation}
    \label{eq particle stress scales}
    \bar{\mathbf{\Sigma}}=\frac{\mathbf{\Sigma}}{\mu \dot{\gamma}} \,, \quad \bar{\mathbf{\Sigma}}^0=\frac{\mathbf{\Sigma}^0}{\mu \dot{\gamma}}\,, \quad \bar{\mathbf{\Sigma}}^p=\frac{\mathbf{\Sigma}^p}{\sigma_{eq}/a} \,.
\end{equation}
Even though both dispersed and suspending fluids are assumed Newtonian, experimental, theoretical and numerical simulations show that emulsions can exhibit non-Newtonian response, including shear-thinning and finite normal stress differences ($N_1>0$ and $N_2<0$). The surface-tension-driven recovery or relaxation of the perturbed drop shape to the minimum surface area for a fixed volume underlies the origin of viscoelasticity. The ratio of this shape relaxation time $\sim \mu a/\sigma_{eq}$ to imposed flow rates $\dot{\gamma}^{-1}$ is a dimensionless group, defined as $Ca$ in Eq. \refeq{eq Ca number} that evokes the Weissenberg number, $Wi$ in elastic soft materials, and likewise, a nonlinear response is observed for $Wi$ > 1. Equations \refeq{eq average stress suspension}-\refeq{eq effective viscosity} hold for the analysis of dilute to concentrated suspensions. Polydispersity in drop sizes of emulsions can be included in the derivation of Eq.~\refeq{eq average stress suspension} if the distribution of drop sizes is known.  At higher concentrations, near the maximum volume fraction of drops, more elaborate constitutive equations are needed to adequately capture the rheological response of the emulsions. The stress and flow behavior of the jammed dense emulsions are discussed in Section \ref{sec: dense}.

\section{Dilute emulsions: small deformation theory and constitutive models} \label{sec: dilute}

In this section, we summarize key features of theoretical and numerical investigations of single-drop dynamics and rheology of dilute emulsions by including three cases: clean drops, surfactant-covered drops, and drops with slip at interfaces. We revisit significant theoretical advances made analytically in the two asymptotic limits of small or large droplet deformations in viscometric flows \cite{rallison1984deformation,stone1994dynamics}. We mention numerical studies used for bridging the gap between the two asymptotic limits for clean drops \cite{rallison1978numerical,cristini2001adaptive,kennedy1994motion}, surfactant-covered droplets \cite{stone1990effects,milliken1993effect,eggleton1998adsorption,eggleton2001tip,lowengrub2004surfactant,lee2006effect,XuLowengrub2006level,tryggvason2008front,pimenta2021study}, and drops with viscous interfaces \cite{pozrikidis1994effects,luo2019influence,gounley2016influence,singh2020deformation,herrada2022stability,singh2023impact}. The approaches discussed here form the starting point for investigations on emulsions containing interfaces with non-Newtonian interfacial rheology, or composed of dispersed or suspending fluid with non-Newtonian rheology response. For example, proteins or particles as emulsifiers lead to interfacial viscoelasticity \cite{williams1997behaviour} or interfacial yield stress, and the presence of lipid membranes and protein gel networks at interface create bending and elastic moduli manifested in suspensions of vesicles and cells including blood. We recommend recent reviews and papers for discussions of emulsions containing complex interfaces that exhibit non-Newtonian interfacial rheology or emulsions formed by using a non-Newtonian dispersed or suspending fluid \cite{guido2011shear, langevin2020emulsions, minale2010models}.

\subsection{Small deformation of drops in shear and extensional flows}
\textbf{Taylor's deformation parameter} In 1932, Taylor generalized Einstein's formula for viscosity of a dilute suspension of hard spheres to derive an expression for the viscosity of dilute emulsions in the limit of low $Ca$, clean interface droplets, and for cases with Newtonian dispersed and suspending fluids. The expression for relative viscosity, $\eta_r$=($\eta$/$\mu$) in the limit of low shear rate (or low Ca) is given by
\begin{equation}
\label{eq Taylor emulsion viscosity}
    \eta_{r}=1+\frac{5}{2}\phi\frac{(\lambda+2/5)}{(\lambda+1)}\ = 1+\frac{5}{2}\phi g_T(\lambda)\,,
\end{equation}
where $g_T(\lambda)={(\lambda+2/5)}/{(\lambda+1)}$ is Taylor's viscosity factor. Recalling that the specific viscosity $\eta_{sp}$=($\eta_r$-1) equals the ratio of the contribution of dispersed and suspending phases to viscosity we deduce, from Taylor's expression, an alternative form for specific viscosity of emulsions, 
\begin{equation}
\label{eq Taylor emulsion specific viscosity}
       \eta_{sp}/{\phi}= {\mathbf{\Sigma}^p_{12}}/{\mu{\dot{\gamma}}}=\frac{5}{2}{g_T(\lambda)}\,.
\end{equation}
In the limit of large $\lambda$, Taylor's viscosity factor goes to unity or $g_T (\lambda)=1$ recovering Einstein's formula for suspensions.  Upon defining the specific viscosity as $\eta_{sp}$=($\eta_r$-1), Eq.~\refeq{eq Taylor emulsion viscosity} yields $\eta_{sp}=(5/2)\,\phi{g_T(\lambda})$.
In the limit of vanishingly small $\lambda$, the parameter $g_T(\lambda)\to 2/5$, implying the relative viscosity of bubbly fluid is just $\eta_{r}$ = 1+$\phi$ and the specific viscosity of bubbly fluid is $\eta_{sp}$ = $\phi$. 

\begin{figure*}[h!]
	\centering
		\includegraphics[scale=.5]{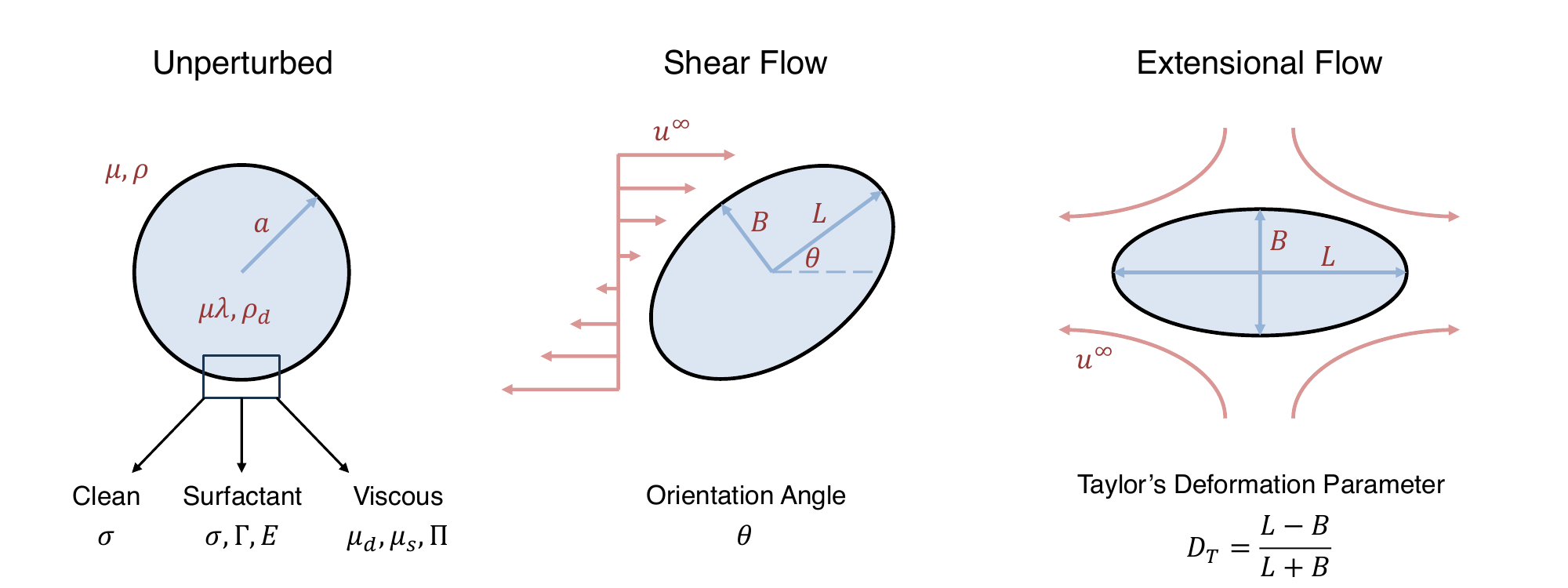}
	\caption{Representative drop deformation in shear and extensional flows; unperturbed shape added as a reference including interfacial properties.}
	\label{FIG:2}
\end{figure*}

Taylor \cite{taylor1932viscosity,taylor1934formation}  was the first to theoretically and experimentally study the deformation of a neutrally buoyant viscous drop in response to imposed shear or extensional flows, and describe how bulk rheology is informed by drop deformation and orientation at the microscopic scale. For a weakly perturbed spherical drop, the shape change can be measured using a scalar quantity called Taylor's deformation parameter defined as
\begin{equation}
\label{eq Taylor Deformation param}
    D_{T}=\frac{L-B}{L+B}\,,
\end{equation}
where $L$ and $B$ are the major and minor axes of the ellipsoid projected onto the velocity-shear rate plane, as shown in Figure \ref{FIG:2}. For large deformations, especially those encountered in response to extensional flows, $L/B$ is usually used instead of $D_{T}$ \cite{stone1994dynamics}. According to Taylor, at steady state, the small deformation of a clean droplet in response to weak shear flows \cite{taylor1934formation}, exhibits
\begin{equation}
    \label{eq D Taylor}
    D_T = \frac{(19 \lambda +16)}{(16 \lambda + 16)} Ca + O(Ca^2) =d_T(\lambda) Ca + O(Ca^2)\,.
\end{equation}
Here, for brevity's sake, we define the viscosity ratio dependent prefactor as the Taylor's deformation prefactor ${d_T}={(19\lambda+16)}/{(16\lambda+16)}$.  


\textbf{Inclination or orientation of deformed drops} In flows with a rotational component of velocity including viscometric shear flows, the ellipsoidal drop (or ellipsoid projection of the deformed drop) orients. An inclination angle, $\theta$ can be measured between the major axis of deformation and the flow direction, as shown in Figure \ref{FIG:2}. Chaffey and Brenner \cite{chaffey1967second} computed the inclination angle exhibited by perturbed drops in weak flows by carrying out perturbation analysis up to second order in $Ca$, leading to the following expression
\begin{equation}
    \label{eq angle Chaffey Brenner}
    \theta = \frac{\pi}{4}-\frac{(2\lambda +3)}{5} \frac{(19 \lambda +16)}{(16 \lambda + 16)}Ca + O(Ca^2)\,,
\end{equation}
or
\begin{equation}
    \label{eq angle Chaffey Brenner 2}
    \theta = \frac{\pi}{4}-\frac{d_{T}}{c_{0}}Ca + O(Ca^2)\,,
\end{equation}
where $c_0(\lambda)=5/(2\lambda+3)$. In dilute emulsions, the flow-induced droplet dynamics depend on the physicochemical properties of the two liquids (density and viscosity), composition-dependent properties of the interface (interfacial tension, interfacial rheology, and surface forces), and the strength and type of imposed flow fields (shear and extensional). Qualitatively, the extend of drop deformation and orientation for clean droplets is influenced by an interplay of viscous and capillary stresses dependent on $Ca$, appropriately defined  by accounting for interfacial tension, deformation rate, and viscosity ratio $\lambda$ ranging 0 to $\infty$.   

\textbf{Clean droplet dynamics} 
In weak flows, $Ca \ll 1$, steady shapes are nearly spherical and the inclination angle $\theta \sim 45^\circ$, to leading order in $Ca$, as sketched in Figure \ref{FIG:2} and first analyzed and visualized by Taylor \cite{taylor1934formation}. For the two extreme values of $\lambda$ i.e., 0 and $\infty$, the deformation prefactor, ${d_{T}}$ ranges between 1 and 1.187, implying that in weak flows, drop deformation parameter ${D_T}$ is linearly dependent on $Ca$. The inclination, $\theta$ according to Chaffey and Brenner equation \refeq{eq angle Chaffey Brenner 2} shows a linear dependence on $Ca$. Visualizing drop deformation under mild flow provides a means of measuring interfacial tension, even when the interfacial tension is extremely small, for example, in water-in-water emulsions or in coacervates. Alternatively, the relaxation of a perturbed drop to its unperturbed state after cessation of flow can be used to measure shape relaxation time and interfacial tension. At higher flow strengths, for a given $\lambda$, droplet shapes become more elongated as $Ca$ increases and the major axis of deformation aligns with the flow direction as the droplet rotates in response to the local vorticity of the flow. In this limit, drops with viscosities below a critical value $\lambda_c \sim 4$ may undergo breakup at a critical flow strength $Ca_c$, whereas high-viscosity drops remain stable for $\lambda>\lambda_c$, for arbitrary $Ca$ \cite{rallison1984deformation,stone1994dynamics}. For example, clean droplets with the same viscosity as the suspending medium undergo breakup at a critical value $Ca_c \approx 0.43$ \cite{cristini2003drop}.

\begin{figure*}[h!]
	\centering
		\includegraphics[scale=.65]{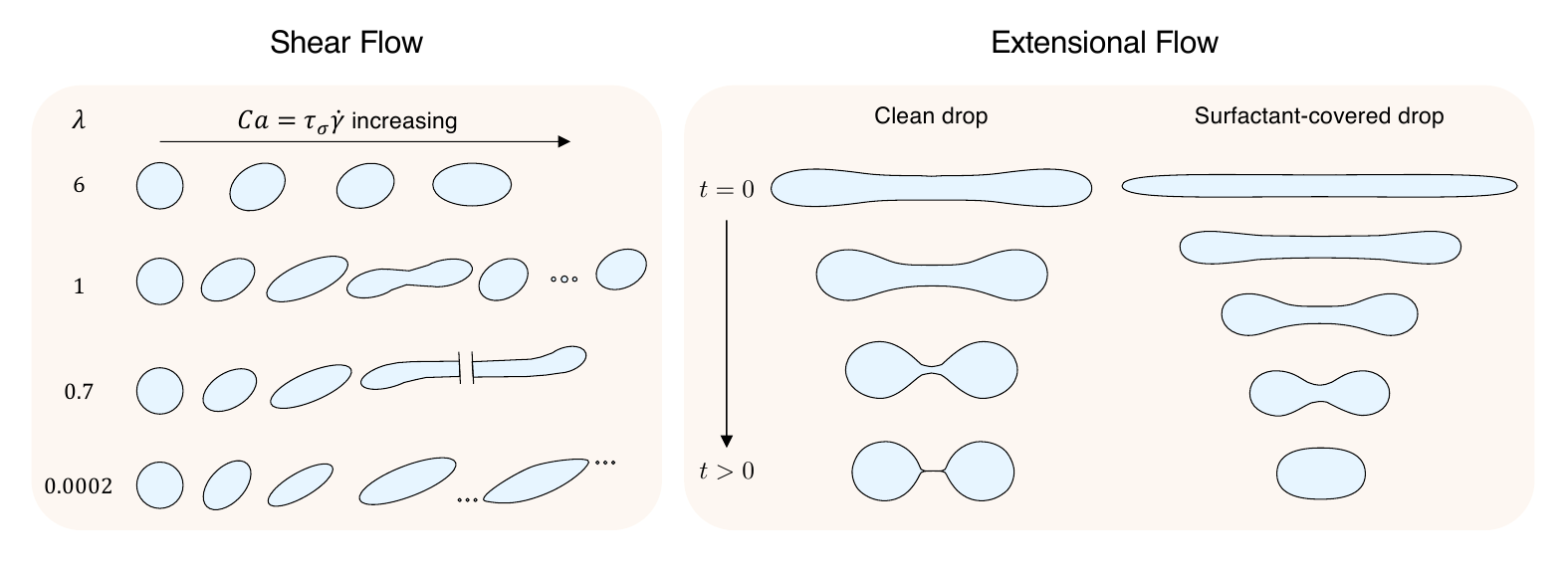}
     \put(-305,18){$(a)$}
     \put(-30,18){$(b)$}
     \caption{Schematic diagram of droplet deformation in shear and extensional flows. Image adapted from Ref.~\cite{rumscheidt1961particle} for shear flow experiments (a) and from Fig.~9 in Ref.~\cite{milliken1993effect} for numerical results in extensional flows (b). The two sets in extensional flows depict snapshots of drop relaxation of clean and surfactant-covered droplets at different dimensionless times as indicated. Details on the experimental data sets in part (a) are listed in Appendix \ref{Appendix: dataset fig 3}.}
	\label{FIG:3}
\end{figure*}

Experiments by Mason and coworkers \cite{rumscheidt1961particle,torza1972particle} characterized the transient and equilibrium drop shapes for  $\lambda < \lambda_c$ and observed breakup modes for clean droplets as illustrated by cases reproduced in Fig.\ref{FIG:3}(a). Breakup modes were observed to depend on a balance between the rate of increase of capillary number up to and across $Ca_c$ and the shape relaxation time. For $\lambda < 0.2$ and high $Ca$ rates, the droplets experience tip-streaming breakup mode; whereas for low enough $Ca$ rates, tip-streaming breakup may be suppressed and the droplet deforms into a thin-liquid thread and breakup into smaller droplets by Rayleigh instability. However, numerical and experimental results in extensional flows support the assumption that tip-streaming instabilities occur only in the presence of surfactants \cite{bentley1986experimental,deBruijn1993tipstreaming,eggleton2001tip,herrada2022stability}. Theoretical and numerical analysis on tip-streaming breakup instability remains an active area of research.


In weak extensional flows, clean droplets attain a stable, stationary shape for all $\lambda$, where the droplet principal axis of deformation is aligned with the flow direction of maximum extension, as illustrated in Fig.~\ref{FIG:3}(b) adapted from Milliken et al. \cite{milliken1993effect}. Here, the transient approach to steady shapes is monotonic, since the flow is vorticity-free. For $Ca=O(1)$, two main regimes of droplet steady deformation are of interest: (i) nearly ellipsoidal shapes are observed for moderate and large $\lambda$, (ii) for $\lambda \lesssim 0.1$, droplets deform into shapes with nearly-pointed ends. For larger values of $Ca$, high-viscosity drops deform into slender threads that eventually breakup into smaller droplets. Low-viscosity drops are able to sustain highly elongated shapes for even larger flow strengths, but will breakup into small droplets via Rayleigh-Plateau instability if $Ca\gg Ca_c$. Drop relaxation after the flow field is switched off may also lead to drop breakup into a chain of droplets of uniform size if the droplet initial shape is sufficiently elongated by the flow.  


\sloppy
\textbf{Deformation of coated droplets} The presence of surface inclusions (e.g., surfactant molecules, proteins, lipids) alters the classical dynamics of transient and steady shapes of clean droplets \cite{fischer2007emulsion}. For surfactant-covered drops, deviations from the clean droplet deformation are governed by a balance among (i) interface convection of surfactants towards regions of high curvature and stagnation points lowering surface tension locally, (ii) local surfactant dilution due to drop deformation and creation of surface area, and (iii) diffusion of surfactant which tends to homogenize the surfactant distribution along the interface. Gradients in surface tension induce Marangoni stresses which act against surface deformation \cite{flumerfelt1980effects, stone1990effects, narsimhan2019shape}. The critical $Ca_c$ for the onset of unsteady deformation and breakup is usually larger compared to clean droplet results, but it can be smaller depending on flow strength and on the local vorticity of the flow \cite{li1997effect}.

Figure \ref{FIG:3}(b) shows the relaxation of clean and surfactant-covered droplets at different times after being initially deformed by an extensional flow. Surfactant redistribution along the droplet surface stabilizes the shape against transient configurations that may lead to droplet breakup. The qualitative behavior of droplets with viscous interfaces in linear flows introduces an additional surface viscous stress to the force balance Eq.~\refeq{eq BC traction jump}, where droplet shape and rheology depend on flow type and emulsion's composition, for example, the relative contribution of shear and dilatational surface viscosities and their relation to surface pressure and surface tension \cite{gounley2016influence,luo2019influence,narsimhan2019shape,singh2023impact}. 
Thus, droplets may attain steady shapes or undergo transient flow-induced deformation, possibly leading to interfacial instabilities and breakup (e.g., tip-streaming, burst, and thread breakup by Rayleigh instabilities) \cite{torza1972particle,grace1982dispersion, rallison1984deformation, stone1994dynamics}. 

\textbf{Deformation a drop in an emulsion with non-Newtonian component(s)} The deformation of drops in emulsions containing one non-Newtonian phase have received more attention than cases with both non-Newtonian phases. Both experiments, analytical theory and numerical simulations show that viscoelasticity changes both the drop deformation and ${Ca_c}$ above which elongated drops breakup. A detailed understanding of interplay of capillarity and viscoelasticity is absent even for drop coalescence and breakup as it requires examination and understanding of the influence of rate-dependent shear viscosity, transient and steady extensional viscosity, normal stress differences, and for polymeric fluids, finite extensibility and non-Hookean elasticity \cite{dinic2017pinch, dinic2019macromolecular, mukherjee2009effects, acrivos1978deformation, ramaswamy1999deformation, ramaswamy1999VEdropdeformation, guido2011shear, greco2002drop}. The experimentalists often utilize constant-viscosity elastic fluids called Boger fluids to isolate the effect of elasticity without interference from rate-dependent viscosity behavior. The constant viscosity elastic fluids can be considered analogs of Oldroyd-B fluids especially while modeling shear rheology response, and for the small deformation case, the simpler second-order fluid model provides a reasonable starting point \cite{greco2002drop,minale2010models}. However, these models are suitable for comparisons with experiments only for slow flows, but the practical problems of dilute and nondilute emulsions often require an understanding of response to strong flows and coupling between rate-dependent viscous and nonlinear viscoelastic effects \cite{greco2002drop, minale2010models, guido2011shear}.

We recommend the classical papers by Leal's group for a comprehensive survey of clean droplet dynamics in unbounded shear and extensional flows \cite{bentley1986experimental,stone1986experimental}, including studies when drop or suspending fluid is non-Newtonian \cite{milliken1991deformation,ramaswamy1999deformation, ramaswamy1999VEdropdeformation}, and direct the interested readers to Guido's review on droplet deformation in confined flows and viscoelastic fluids \cite{guido2011shear}. Numerical methods used for visualization of drop deformation in nondilute emulsions for different choices of $\lambda$, flow types, $Ca$, interfacial properties, and Newtonian or viscoelastic fluids are summarized in a later section. Next we describe how drop deformation, orientation and shape relaxation provide ingredients for deriving or prescribing constitutive models for emulsions by incorporating the role of interfacial and bulk properties of dispersed and suspending fluids. 



\subsection{Constitutive models for dilute emulsions}
\textbf{Evolution of the shape distortion tensor} In the limit when a suspended neutrally buoyant, clean droplet of undeformed, spherical radius $a$ deviates from sphericity to slight ellipticity, the perturbed drop shape is described by \cite{frankel1970constitutive,cox1969deformation,rallison1980note}
\begin{equation}
    \label{eq drop surface}
    S(t)=r(t)-a \left(1+ \epsilon \frac{ \bx \cdot \mathbf{A}(t) \cdot \bx }{r^2}\right) + O(\epsilon^2)=0\,,
\end{equation}
where $\epsilon \ll 1$ is a perturbation parameter,  and $r=(\bx \cdot \bx )^{1/2}$. The shape distortion tensor, $\mathbf{A}$, measures the droplet deformation embedded in the current configuration relative to a reference configuration. The components of $\mathbf{A}$, a second-order tensor, are determined in terms of a second-order deformation gradient tensor, $\mathbf{F}$, that maps the deformation of the material lines from reference to current configuration \cite{spencer2004continuum}. Please refer to Appendix \ref{appendix: small deformation theory} for a more detailed description.

The rate of change in the droplet shape depends on the kinematics of the imposed flow $\mathbf{u}^{\infty}=(\mathbf{E}+\mathbf{W})\cdot \bx$, and thus on the rate-of-strain tensor $\mathbf{E}$ and vorticity tensor $\mathbf{W}$. The distortion tensor $\mathbf{A}$ can be used to calculate the Taylor deformation parameter ${D_T}$, orientation (inclination angle in shear flows), and define rheological material properties of the emulsion. The evolution of the distortion tensor in a reference frame that translates and rotates with the droplet \cite{rallison1984deformation,oliveira2015emulsion} is captured by the following expression
\begin{eqnarray}
    \label{eq evolution distortion}
    \epsilon \frac{\partial \mathbf{A}}{\partial \bar{t}}-Ca \bar{\mathbf{W}}\cdot \epsilon  \mathbf{A}+\epsilon Ca  \mathbf{A}\cdot \bar{\mathbf{W}} =  Ca \,c_0(\lambda) \bar{\bE} \nonumber&\\-  c_1(\lambda) \epsilon \mathbf{A} +O(\epsilon Ca, \epsilon^2)\,.
\end{eqnarray}
Here the two coefficients $c_0(\lambda)=5/(2 \lambda +3)$ and $c_1(\lambda)=40(\lambda +1)/[(19 \lambda +16)(2 \lambda+3)]$ depend primarily on viscosity ratio, $\lambda$. The coefficient $c_0(\lambda)$ appeared in the definition of the deformation parameter, $D_T$ in Eq.~\refeq{eq angle Chaffey Brenner 2}. The dimensionless quantities are defined for time, strain rate tensor and vorticity tensor, respectively as $\bar t = t \, /(\mu a/\sigma)$, $\bar{\bE} = \bE/\dot{\gamma}$,$\bar{\mathbf{W}} = \mathbf{W}/\dot{\gamma}$, and $\vert \mathbf{A} \vert =1$. A derivation of Eq.~\refeq{eq evolution distortion} is included in the Appendix \ref{appendix: small deformation theory} for completeness. The right-hand side of Eq.~\refeq{eq evolution distortion} captures how the rate of change of $\mathbf{A}$ is contributed by two competing terms. The first term distorts away from spherical shape and is linearly dependent on the rate of strain, whereas the second term restores unperturbed shape and depends on $\mathbf{A}$. The neglected terms of $O(\epsilon^2)$  correspond to harmonics higher than second, whereas terms of $O(\epsilon Ca)$ arise from the straining flow acting on the distorted shape \cite{rallison1984deformation}.  It is possible to rewrite $c_1(\lambda)=1/[2{c_0}{d_T}]$ and reframe the Eq. \refeq{eq evolution distortion} to appear as:
\begin{eqnarray}
    \label{eq evolution distortion d_T}
    \epsilon \frac{\partial \mathbf{A}}{\partial \bar{t}}-Ca \bar{\mathbf{W}}\cdot \epsilon  \mathbf{A}+\epsilon Ca  \mathbf{A}\cdot \bar{\mathbf{W}} =  Ca \,c_0 \bar{\bE} \nonumber&\\-  \frac{1}{2c_{0}d_{T}} \epsilon \mathbf{A} +O(\epsilon Ca, \epsilon^2)\,.
\end{eqnarray}
The form of Eq.~\refeq{eq evolution distortion} or Eq.~\refeq{eq evolution distortion d_T} reveals two small deformation regimes: (i) for weak flows (i.e., $\epsilon \sim Ca \ll 1$ and $\lambda=O(1)$), the distortion is limited by strong interfacial tension effect, and (ii) large-$\lambda$ and arbitrary $Ca$ but not too large for flows with sufficient vorticity where  $\epsilon\sim \lambda^{-1} \ll 1$. For a given flow type and small parameter $\epsilon$, Eq.~\refeq{eq evolution distortion} is solved for the distortion tensor $\mathbf{A}$. Here, we summarize up to second-order deformation theories for clean droplet deformation and rheology in viscometric flows and include the results for surfactant-covered drops, interfacially viscous drops, and drops with interfacial slip conditions. The interfacial slip case is key to understanding the rheology of polymer blends as emulsions, which are formed by phase separation.  

\textbf{Clean droplets in shear flows} For a clean droplet in weak shear flows where $\epsilon=Ca \ll 1$ and $\lambda=O(1)$, the deformation parameter ${D_T}$ derived by Taylor \cite{taylor1934formation} and the inclination angle $\theta$ as demonstrated by Chaffey and Brenner \cite{chaffey1967second} shows a linear dependence on $Ca$. These expressions discussed in the previous section and given by Eqs.~\refeq{eq D Taylor} and \refeq{eq angle Chaffey Brenner 2}, respectively, are reproduced here for clarity of presentation,
\begin{equation}
    \label{eq D Taylor 2}
    D_T = \frac{19 \lambda +16}{16 \lambda + 16} Ca + O(Ca^2)={d_T}Ca + O(Ca^2)\,,
\end{equation}
\begin{equation}
    \label{eq angle Chaffey Brenner 3}
    \theta = \frac{\pi}{4}-\frac{d_{T}}{c_0} Ca + O(Ca^2)\,.
\end{equation}
In the other limit when $Ca=O(1)$ and $\epsilon=\lambda^{-1} \ll 1$, the leading order solutions for the Taylor deformation parameter and inclination angle are
\begin{equation}
    \label{eq large lambda Taylor}
    D_T=\frac{5}{4} \lambda^{-1} +O(\lambda^{-2})\,, \quad \theta=\frac{10}{19} \frac{\lambda^{-1}}{Ca} +O(\lambda^{-2})  \,.
\end{equation}
Higher-order theories have been developed; for detailed derivation and formulas, see Refs.~\cite{barthes1973deformation,rallison1980note,vlahovska2009small,oliveira2015emulsion,narsimhan2019shape}. 

 For clean drops in shear flows in the weak flow limit when $\epsilon=Ca \ll 1$ and arbitrary $\lambda$, a second-order deformation analysis \cite{schowalter1968rheological,barthes1973rheology,vlahovska2009small} leads to the following equations that describe the characteristic rheological behavior of dilute emulsions
\begin{equation}
    \label{eq shear stress 12 clean}
    \frac{\Sigma^p_{12}}{\mu \dot{\gamma}}=\frac{5}{2} {g_T}- \frac{4}{5}{d_T} D_1 (\lambda) Ca^2  +O(Ca^3) \,,
\end{equation}
\begin{equation}
    \label{eq shear N1 clean}
    \frac{N^p_1}{\mu \dot{\gamma}}=\frac{32}{5} {d_T}^2 Ca\,, 
\end{equation}
\begin{equation}
    \label{eq shear N2 clean}
    \frac{N^p_2}{\mu \dot{\gamma}}=-\frac{1}{2} \frac{N^p_1}{\mu \dot{\gamma}} -\frac{4}{5}Ca \, {d_T} D_{2}(\lambda)
\end{equation}
where the coefficients $D_0$ and $D_1$ are listed in Appendix \ref{appendix: small deformation coeffs}. Equations \refeq{eq shear stress 12 clean}-\refeq{eq shear N2 clean} reveal the characteristic shear-thinning behavior of emulsion flows with finite positive and negative first and second normal stress differences, respectively. 

Since ${\mathbf{\Sigma}^p_{12}}/{\mu{\dot{\gamma}}}=\eta_{sp}/{\phi}$, the explicit use of ${g_T}$, ${d_T}$ and ${c_0}$ in writing the contributions of dispersed phase (or particulates) to shear stress, specific viscosity, and normal stress differences provides two key benefits. The expressions appear more compact and allow for clearer comparisons. Also, from $\lambda$ ranging from zero (bubble) to infinity (rigid particle), ${d_T}$ varies from 1 to 1.187, whereas ${g_T}={(\lambda+2/5)}/{(\lambda+1)}$ increases from 2/5 to 1.

In the limit when $\epsilon=\lambda^{-1} \ll 1$ for arbitrary $\lambda Ca$, Oliveira \& da Cunha \cite{oliveira2015emulsion} developed a second-order perturbation theory in powers of $\lambda^{-1}$ and showed that 
\begin{equation}
    \label{eq shear stress 12 clean lambda}
    \frac{\Sigma^p_{12}}{\mu \dot{\gamma}}=\left(\frac{5}{2}-\frac{25}{4 \lambda}\right) + \frac{5}{\lambda} \frac{20/19}{\left[(20/19)^2+(\lambda Ca)^2\right]} \,,
\end{equation}
\begin{equation}
    \label{eq shear N1 N2 clean lambda}
    \frac{N^p_1}{\mu \dot{\gamma}}=\frac{10}{\lambda} \frac{(\lambda\,Ca)^2}{\left[(20/19)^2+(\lambda Ca)^2\right]}\,, \quad  \frac{N^p_2}{\mu \dot{\gamma}}=-\frac{29}{133} \frac{N^p_1}{\mu \dot{\gamma}} \,.
\end{equation}
The shear rheology of high-viscosity drops reveals two limits. When $Ca\ll 1$ or weak flows, emulsions of high-viscosity drops behave as Boger fluids with shear rate independent viscosity and vanishing, but finite normal stress differences; a similar behavior is observed for $Ca=O(1)$.

\textbf{Surfactant-covered drops} Vlahovska et al. \cite{vlahovska2009small} extended the small-deformation theory for clean droplets to surfactant-covered drops valid for arbitrary viscosity ratios and elasticity parameter. In weak flows, the deformation and inclination angle at leading order are
\begin{equation}
    \label{eq D Taylor surfactant}
    D_T=  \frac{5}{4} Ca + O(Ca^3) \,,
\end{equation}
and
\begin{equation}
    \label{eq angle surfactant}
    \theta =  \frac{\pi}{4} - \left[\frac{(32+23 \lambda)\beta + 4+\lambda}{48 \beta}\right] Ca + O(Ca^2) \,.
\end{equation}
In weak flows free of vorticity, the stationary shape and surfactant distribution are independent of viscosity ratio since Marangoni stresses immobilize the droplet interface \cite{milliken1993effect,vlahovska2009small}. The rheological material functions for drops covered with insoluble surfactants in shear flow for surface elasticity parameter $\beta=CaMa$ are
\begin{equation}
    \label{eq shear stress 12 surfactants}
    \frac{\Sigma^p_{12}}{\mu \dot{\gamma}}=\frac{5}{2} - D_{3}(\lambda,\beta)  Ca^2  +O(Ca^3) \,,
\end{equation}
\begin{equation}
    \label{eq shear N1 N2 surfactants}
    \frac{N^p_1}{\mu \dot{\gamma}}=\frac{5}{2} \frac{(4 \beta+1)}{\beta} Ca\,, \quad \frac{N^p_2}{\mu \dot{\gamma}}=-\frac{1}{2} \frac{N^p_1}{\mu \dot{\gamma}} + \frac{75}{28} Ca\,,
\end{equation}
where the coefficient $D_3$ is defined in Appendix \ref{appendix: small deformation coeffs}
Note that, in the limit of $Ca \to 0$,  inserting Eq.~\refeq{eq shear stress 12 surfactants} into Eq.~\refeq{eq effective viscosity} yields Einstein's classical result $1+(5/2) \phi$ given by Eq.~\refeq{eq Taylor emulsion viscosity} with $g_T (\lambda)=1$ and emulsion rheology follows the behavior of a suspension of rigid spheres with vanishing normal stress differences. 

Recently, Narsimhan \cite{narsimhan2019shape} developed a higher order small deformation theory for shape and rheology of drops covered with viscous interfaces expanding from previous classical works by Oldroyd \cite{oldroyd1955effect} and Flumerfelt \cite{flumerfelt1980effects}. To leading order, in the limit as $\epsilon=Ca \ll 1$ and $\lambda$, $Bq_s$,$Bq_d \sim O(1)$
\begin{equation}
    \label{D Taylor viscous}
    D_T= \frac{1}{2} \alpha_0 Ca  \,, \quad \alpha_0=\frac{1}{8}\frac{19 \lambda + 16+24 Bq_d +8 Bq_s}{\lambda^* +1} 
\end{equation}
$\alpha_0$ is the Taylor deformation parameter, $\lambda^*=\lambda +(6/5)Bq_d+(4/5)Bq_s$ is a modified viscosity ratio, and the inclination reduces to 
\begin{equation}
    \label{eq inclination angle viscous Ca}
    \theta=\frac{\pi}{4}+\frac{Ca}{2}a^{-1}_D\,,
\end{equation}
where $a_D(\lambda,Bq_s,Bq_d)$ is an expansion coefficient \cite{narsimhan2019shape} defined in Appendix \ref{appendix: viscous interface}.
The corresponding analytical formulas for shear rheology are
\begin{equation}
    \label{eq viscosity viscous and N1 Ca}
    \frac{\Sigma^p_{12}}{\mu \dot{\gamma}}=\frac{5}{2}{g_T} \,, 
\end{equation}
\begin{equation}
    \label{eq viscous N1 Ca}
    \frac{N^p_1}{\mu \dot{\gamma}}=\frac{8}{5} \alpha^2_0\,,
\end{equation}
\begin{equation}
    \label{eq viscous N2 Ca}
    \frac{N^p_2}{\mu \dot{\gamma}}=-\frac{1}{2}\frac{N^p_1}{\mu \dot{\gamma}} + \frac{3 \alpha_0}{70} \frac{(25 \lambda^{*^{2}}+41 \lambda +24 Bq_d+4)}{(\lambda^*+1)^2}\,,
\end{equation}
where shear-thinning effects are $O(Ca^2)$ contributions \cite{narsimhan2019shape}.

In the other small deformation limit when $\epsilon \ll 1$ and $Ca = O(1)$,
\begin{equation}
    \label{D Taylor viscous 2}
    D_T= \frac{1}{2} \hat{a}_E (1+\hat{a}_E)+O(\epsilon^3)  \,, \quad \theta=-\frac{1}{2}\frac{\hat{a}_D}{Ca}+O(\epsilon^2) \,,
\end{equation}
where the small parameter $\epsilon=\lambda^{-1}$ or $Bq_s^{-1}$ for $Bq_s\sim Bq_d$. The form of the coefficients $\hat{a}_D$ and $\hat{a}_E$ are shown in the Appendix \ref{appendix: viscous interface}. In this limit, small-deformation theory indicates that the emulsions of highly viscous internal or surface viscosities behave approximately as rigid spheres with no shear-thinning and no significant elastic effects. This observation is in agreement with the small-deformation theory for high-viscosity drops in weak flows \cite{vlahovska2009small,oliveira2015emulsion}. 

\textbf{Droplets with slip at the interfaces} Ramanchandran \& Leal \cite{ramachandran2012effect} developed a second-order small deformation analysis for drops with interfacial slip in weak flows. The model captures the anomalous decrease in relative viscosity measured in emulsions formed by immiscible polymer blends.  The viscometric functions in shear flow are

\begin{equation}
    \label{eq viscosity slip}
     \frac{\Sigma^p_{12}}{\mu \dot{\gamma}}=  \frac{5 \lambda(2\bar \alpha+1)+2}{2 \lambda(5\bar \alpha+1)+2}+O(Ca^2) \,,
\end{equation}

\begin{equation}
    \label{eq viscosity slip 2}
     \frac{\Sigma^p_{12}}{\mu \dot{\gamma}}=  \frac{(5/2) g_T+h(\lambda,\bar \alpha)}{1+h(\lambda,\bar \alpha)}+O(Ca^2) \,,
\end{equation}

\begin{equation}
    \label{eq shear N1 N2 slip}
    \frac{N^p_1}{\mu \dot{\gamma}}= f(\lambda,\bar \alpha) Ca  \,, \quad \frac{N^p_2}{\mu \dot{\gamma}}=\left[\frac{g(\lambda,\bar \alpha)}{4}-\frac{f(\lambda,\bar \alpha)}{2}\right]Ca\,,
\end{equation}
where $h(\lambda,\bar \alpha)=5 \lambda \bar \alpha/(\lambda+1)$, and the functions $f$ and $g$ are defined in Appendix \ref{appendix: slip interface}, for completeness \cite{ramachandran2012effect}.
In extensional uniaxial flow, the theory predicts
\begin{equation}
    \label{eq D Taylor slip}
    \frac{\tilde \mu/\mu-3}{\phi} =\frac{5 \lambda(2\bar \alpha+1)+2}{2 \lambda(5\bar \alpha+1)+2} + \frac{g(\lambda,\bar \alpha)}{4} Ca +O(Ca^2) \,,
\end{equation}

\begin{equation}
    \label{eq D Taylor slip 2}
    \frac{\tilde \mu/\mu-3}{\phi} =\frac{(5/2) g_T+h(\lambda,\bar \alpha)}{1+h(\lambda,\bar \alpha)} + \frac{g(\lambda,\bar \alpha)}{4} Ca +O(Ca^2) \,,
\end{equation}
where $\tilde \mu=3 \mu$ is the Trouton viscosity for the pure suspending fluid ($\phi=0$), and
\begin{equation}
    \label{eq extension slip}
    \frac{\tilde \mu}{\mu}=\frac{\Sigma^p_{33}-\Sigma^p_{11}}{\mu \dot{\gamma}}=\frac{\Sigma^p_{33}-\Sigma^p_{22}}{\mu \dot{\gamma}}\,,
\end{equation}
by definition. The effect of interfacial slip on material functions in shear and extensional flows is more pronounced for values of viscosity ratio $\lambda > O(1)$. Slip has a stronger influence  in response to extensional flows than shear. The analytical results indicate that slip hinders droplet deformation and decrease effective viscosity of the emulsion. However, quantitative agreement between theory and experiments is not verified even in the limit of infinite slip, suggesting that additional physical mechanisms might contribute to the pronounced viscosity reduction observed in experiments \cite{utracki1983melt}.

In this section on rheology of dilute emulsions, wall effects and interparticle interactions were neglected. The material functions (specific viscosity and normal stresses) for clean as well as complex interfaces show characteristics of non-Newtonian response, including a nonlinear relation between stresses and the rate of strain giving rise to fluid memory. Frankel and Acrivos \cite{frankel1970constitutive} extended the works by Chaffey \& Brenner \cite{chaffey1967second}, Schowalter, Chaffey \& Brenner \cite{schowalter1968rheological} and, Cox \cite{cox1969deformation} and proposed a set of constitutive equations that capture the transient effects of droplet deformation induced by an imposed linear time-dependent flow field. Appendix \ref{Appendix: Frankel Acrivos} revisits the Frankel and Acrivos analysis, and retraces the steps and assumptions that lead them to a three-parameter Jeffreys-like constitutive equation for emulsions of the form 
\begin{equation}
\label{eq constitutive Jeffreys form}
     \displaystyle \Sigma_{ij} + \Lambda \frac{\mathcal{D}\Sigma_{ij}}{\mathcal{D}t} =
- p \delta_{ij}+ \eta\left(\dot{\gamma}_{ij} +\Lambda_J \frac{\mathcal{D}\dot{\gamma}_{ij}} {\mathcal{D}t}\right)\,,
\end{equation}
where we let $\dot{\gamma}_{ij}=2E_{ij}$. The three parameters are viscosity, and two timescales. The material relaxation time is dependent on the shape relaxation time as follows
\begin{eqnarray}
    \label{eq Lambda relaxation time}
    \Lambda = \frac{(2\lambda + 3)(19\lambda + 16)}{40(\lambda + 1)} \tau_\sigma = c_1(\lambda)^{-1}\tau_\sigma \,,
\end{eqnarray}
and a Jeffreys emulsion retardation time can be determined from:
\begin{equation}
    \label{eq Jeffrey retardation time}
    \displaystyle \Lambda_J=\Lambda\left(1-  \frac{4}{5}\frac{\Lambda}{\tau_{\sigma}}\frac{\phi}{\eta_r(\lambda,\phi)} 
    \right)\,.
\end{equation}

\section{Non-dilute emulsions: constitutive models and numerical methods} 
\label{sec: concentrated}
\subsection{Constitutive models based on small deformation and effective medium theories}
Constitutive equations proposed for non-dilute emulsions aim to account for finite effects of drop deformations, interactions, and microstructure with respect to each other at dispersed-phase volume fractions typically above $10\%$. 

\textbf{Oldroyd's effective medium theory (1953)} Oldroyd \cite{oldroyd1953elastic} used an effective medium approach to derive an expression for the effective viscosity of the semi-dilute emulsions following a perturbation analysis proposed by Frölich \& Sack \cite{frohlich1946theory} for suspensions of elastic spheres. Finite size effects of the higher dispersed-phase volume fraction were included using a cell model. The cell model represents a composite system consisting of a drop (or a particle) surrounded by a volume of suspending fluid in a cell beyond which the emulsion (or a suspension) is seen as a continuum material. This condition is enforced by a modified far-field velocity boundary condition for the disturbance flow generated by the particle (here the generalized term particle engulfs all different types, e.g., drops, capsules, vesicles, rigid and deformable particles, and blood cells). Specifically, Eq.~\refeq{BC far field} is evaluated at a truncated far field position $b/a \sim \phi^{-1/3}$, where $b$ is the characteristic size of the cell in which pressure and velocity disturbances are evaluated, and $a$ is the particle size. Oldroyd's effective medium analysis results in the following expression for the effective relative viscosity of an emulsion:
\begin{equation}
    \label{eq Oldroyd}
    \eta_r=1+ \phi \frac{5 \lambda+2}{2(\lambda+1)}\left(1+\phi \frac{\left(5 \lambda +2\right)}{5(\lambda+1)}\right)\,.
\end{equation}
Using Taylor's factor ${g_T}={(\lambda+2/5)}/(\lambda+1)$ and specific viscosity as $\eta_{sp}$=($\eta_r$-1), Oldroyd's relative viscosity relation can be rewritten in an alternative and compact form as
\begin{equation}
    \label{eq Oldroyd specific}
    \frac{\Sigma^p_{12}}{\mu \dot{\gamma}}=\frac{\eta_{sp}}{\phi}= \frac{5}{2}{g_T}\left(1+\phi {g_T}\right)\,.
\end{equation}

\textbf{Choi \& Schowalter's small deformation theory for nondilute emulsions} Choi \& Schowalter \cite{choi1975rheological} proposed an alternative derivation of effective viscosity of nondilute emulsions by expanding on the stress-averaged, small-deformation theories of Frankel \& Acrivos \cite{frankel1970constitutive} and Cox \cite{cox1969deformation}, by accounting for interparticle interactions and higher-order effects of disperse-phase volume fraction. In steady shear flow, Choi \& Schowalter's constitutive equation yields the following expression for relative viscosity of emulsions.
\begin{equation}
    \label{eq Choi Schowalter}
    \eta_r=1+ \phi \frac{5 \lambda+2}{2(\lambda+1)}\left(1+\phi \frac{5}{4}\frac{\left(5 \lambda +2\right)}{(\lambda+1)}+ O(\phi^{5/3})\right)\,,
\end{equation} 
 
On comparing their expressions with Oldroyd's model from 1953, Choi and Schowalter noted that their coefficient for the second order term was 15.6 compared to 2.5 computed by Oldroyd, and the larger coefficient agrees better with the experiments and Oldroyd's expectation. Alternatively, the shear rheology response can be rewritten in terms of scaled shear stress and normal stress differences as follows:
\begin{equation}
    \label{eq Choi Schowalter gT}
    \frac{\Sigma^p_{12}}{\mu \dot{\gamma}}=\frac{\eta_{sp}}{\phi}= \frac{5}{2}{g_T}\left(1+\frac{25}{4}{g_T}\phi+O(\phi^{5/3}) \right)\,,
\end{equation}
\begin{equation}
    \label{eq Choi Schowalter N1 Ca}
    \frac{N^p_1}{\mu \dot{\gamma}}=\frac{32}{5}{d_T}^2 Ca\phi\,,
\end{equation}
\begin{equation}
    \label{eq Choi Schowalter N2 Ca}
    \frac{N^p_2}{N^p_1}
    =   - \frac{1}{7}\frac{29{\lambda}^2+61\lambda+50}{19{\lambda}^2+35\lambda+16}.
\end{equation}
The expressions for the normal stress differences, given by Eqs.~(30) and (31) in Ref.~\cite{choi1975rheological} are presented here in alternative form. The value of ${N^p_2}/{N^p_1}$ goes to, approximately, -0.218 for $\lambda\to \infty$ and -0.446 as $\lambda \to 0$. Furthermore, the Choi-Schowalter model leads to equations for extensional viscosity that reveal a linearly increasing value with strain rate for uniaxial and linearly decreasing value for biaxial extensional flow.

\begin{figure*}[h!]
	\centering
		\includegraphics[scale=.44]{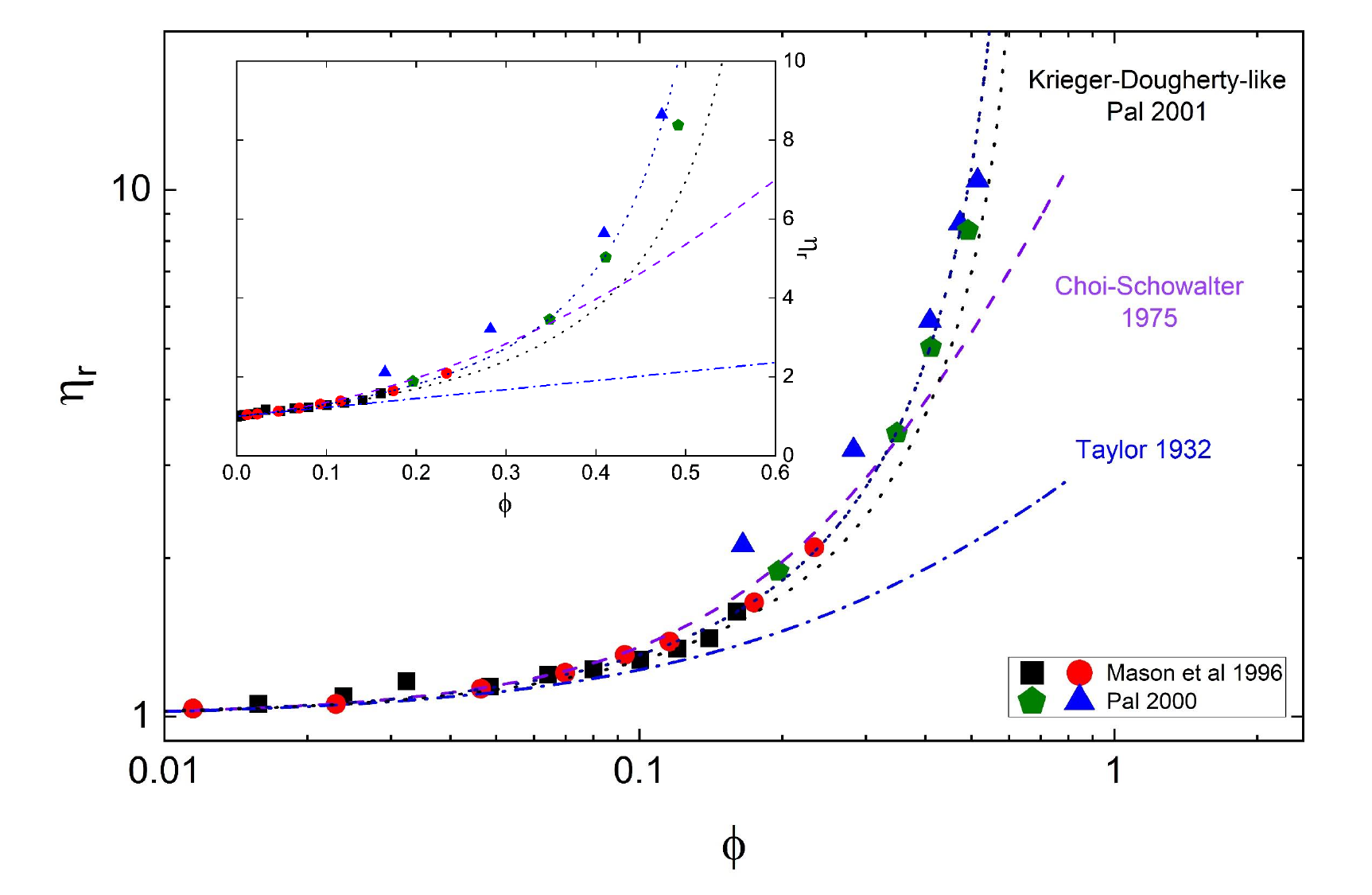}
	\caption{Comparison of theory and empirical relations for effective viscosity models of dilute and concentrated emulsions, respectively. Taylor's effective viscosity relation is obtained by inserting Eq.~\refeq{eq shear stress 12 clean} into Eq.~\refeq{eq effective viscosity} (dash-dotted line), Choi \& Schowalter is given by Eq.~\refeq{eq Choi Schowalter} (dashed line), and Eq.~\refeq{eq Pal effective viscosity II} is used for the Krieger-Dougherty-like curve (dotted lines) where two dotted lines are added: one using data from Ref.~\cite{pal2001novel} (black), and another one with $\lambda$ adjusted to best fit the experimental data (blue). See Appendix \ref{Appendix: dataset fig 3} for details on the datasets and models used.}
	\label{FIG:Effective viscosity comparison}
\end{figure*}

A comparison between the Choi-Schowalter model prediction based on Eq.~\refeq{eq Choi Schowalter} and experimental data is shown in Fig.~\ref{FIG:Effective viscosity comparison}. The striking agreement emphasizes that the model captures the observations relatively well valid up to terms $O(\phi^2)$ in the semi-dilute regime. The use of logarithmic x-axis for volume fraction, $\phi$ helps to observe that the original analysis by Taylor \cite{taylor1932viscosity} captures the linear dependence of $\eta_{r}$ on $\phi$ in the dilute regime reasonably well. Before Choi and Schowalter \cite{choi1975rheological}, Yaron \& Gal-Or \cite{yaron1972viscous} had proposed a similar model considering a free-surface cell approach, that allows for surfactant effects, but without including drop deformation effects. Later generalizations of Frankel \& Acrivos, Oldroyd, and Choi \& Schowalter viscosity models were developed to include non-Newtonian effects of the drop and suspending phases \cite{palierne1990linear,larson1999structure}, though a lot of open questions remain regarding the influence of viscoelasticity of the suspending or dispersed liquid phases and the interface. 

\subsection{Empirical equations} Empirical relations are often used to capture the effective viscosity of emulsions of spherical droplets ($Ca \to 0$) as a function of $\phi$ in analogy with suspensions of rigid spheres. For example, a modification of classical suspension models yields the following equation \cite{yaron1972viscous,choi1975rheological,pal1992rheology}, 
\begin{equation}
    \label{eq Pal effective viscosity}
    \eta_{r}=\exp \left(\frac{5/2\,\, \phi}{1-\phi/\phi_{m}}\right)^\alpha\,,
\end{equation}
where relative viscosity, $\eta_{r}\equiv \eta/\mu$ is the zero-shear-rate viscosity normalized by the viscosity of the suspending medium, 
$\alpha=(2/5 +\lambda)/(1+\lambda)$. Here, $\phi_{m}$ is the emulsion maximum volume fraction at  which the effective viscosity \refeq{eq Pal effective viscosity} diverges. The value of $\phi_m$ decreases with increasing viscosity ratio ranging from $0.63-0.64$ for high-viscosity drops \cite{larson1999structure}. In the dilute regime, $\phi \ll 1$, Eq.~\refeq{eq Pal effective viscosity} reduces to Taylor's result (see Eq.~\refeq{eq shear stress 12 clean}). In the limit when $\lambda \to \infty$ and arbitrary concentrations, Eq.~\refeq{eq Pal effective viscosity} recovers a Krieger-Dougherty-like empirical viscosity relation for suspensions of hard spheres \cite{larson1999structure}.

For finite values of viscosity ratio, an alternative Krieger-Dougherty-like viscosity model is \cite{pal2001novel}
\begin{equation}
    \label{eq Pal effective viscosity II}
    \eta_{r}\left[\frac{2 \eta_r + 5 \lambda}{2 + 5\lambda}\right]^{3/2}=\left(1-\phi/\phi_{m}\right)^{-2.5 \phi_m}\,.
\end{equation}
Predictions for Eq.~\refeq{eq Pal effective viscosity II} compared to experimental data are shown in Fig.~\ref{FIG:Effective viscosity comparison}. The inset shows data plotted on a linear-linear axis. The corresponding plot shown using log-log scale helps to emphasize how well Taylor's pioneering theory \cite{taylor1932viscosity} captures the rheology of dilute emulsions (details about properties of dispersed and suspending liquids are included in Appendix \ref{Appendix: dataset fig 3}). The comparison of theory and experiments reveals that the Choi-Schowalter model \cite{choi1975rheological, larson1999structure} captures the non-linearity introduced by drop-drop interactions in nondilute emulsions, but the impact of higher order interactions and microstructure require a careful consideration for $\phi$ > 0.4 or so. For a comprehensive review on the empirical viscosity models for concentrated emulsions see Ref.~\cite{pal2023recent}. For nondilute emulsions, normal stress differences become important and shear-thinning effects are also observed at higher shear rates\cite{KimMason017advances, foudazi2015physical, langevin2020emulsions}.

\subsection{Doi-Ohta, Maffettone-Minale and other alternatives to small deformation theory}
Constitutive models described so far in this article primarily focus on understanding how drop deformation influences and determines emulsion rheology, starting with Eq.~\refeq{eq average stress suspension}. The small deformation theory works best for vanishingly low $Ca$ values or in flows with drops closed to spherical. The results of the small deformation theories are recovered by a series of phenomenological models, inspired by the Maffettone-Minale model \cite{maffettone1998equation} as briefly reviewed here, that start by assuming that drops are ellipsoidal and allow for drop rotation by vorticity, deformation by strain rate and relaxation to unperturbed state governed by interfacial tension. Slender body theories are used for capturing drop deformation and emulsion rheology for cases when drops are already extended to thread-like shape, usually under high extensional flow, and though transverse cross-section is assumed to remain circular due to surface tension, the drops in forms of long fluid threads stretch and deform along their axial direction \cite{taylor1966conical,acrivos1978deformation, khakhar1986deformation,hinch1980long, herrada2022stability}. The alternative models are needed as emulsions formed by mechanical mixing of immiscible liquids or by phase separation can contain drops of various sizes and shapes, and during flow, these drops can deform, coalesce and breakup leading to complex evolution of drop size and shape distribution. In 1991, Doi and Ohta \cite{doi1991dynamics} proposed an alternative model for emulsion rheology that characterized the rheology in terms of interfacial orientation and area, and three parameters (viscosity, surface tension and volume fraction) but without introducing drop size as an intrinsic length scale. In 1998, Maffettone and Minale \cite{maffettone1998equation} provided an alternative phenomenological model for emulsion rheology by assuming ellipsoidal drops and for small to moderate deformation rates. Several variants of ellipsoidal models have been developed since, including for viscoelastic drops or suspensing fluid, and likewise the Doi-Ohta formalism has been extended to allow for variable viscosity ratios, as summarized in several reviews \cite{minale2010models, larson1999structure, guido2011shear, tucker2002microstructural, cardinaels2016morphology}.

\textbf{The Doi-Ohta model and its descendants} The Doi-Ohta model, originally proposed for equiviscous and equidense concentrated emulsions \cite{doi1991dynamics} adopted a coarse-grained approach to incorporate the influence of change in area and orientation of the interface to the emulsion stress.  The Doi-Ohta model describes the total macroscopic stress for an emulsion as:
\begin{equation}
    \label{eq Doi-Ohta stress}
    \mathbf{\Sigma}=\mathbf{\Sigma}^0 - \sigma\mathbf{q}\ = 2 \mu \langle \bE \rangle-\langle p\rangle \bI - \sigma\mathbf{q}\,,
\end{equation}
\begin{equation}
    \label{eq Doi-Ohta q Q}
 \mathbf{q} =\frac{1}{V} \int_S dS (\mathbf{n}\mathbf{n} - \mathbf{I}),\quad {Q}=\frac{1}{V} \int_S dS\,.
\end{equation}
Here the interface tensor $\mathbf{q}$, a symmetric and second rank tensor, tracks the orientation of the interface where $dS$ is a differential element of interfacial area, V is the system volume, and the integral is over all interfaces. A scalar $Q$ quantifies the amount of total interfacial area per unit volume, and has dimensions of inverse length or 1/length. The interface tensor $\mathbf{q}$ and the parameter $Q$ are related by a pair of evolution differential equations \cite{larson1999structure, doi1991dynamics}. The original Doi-Ohta model was restricted to a concentrated emulsion, obtained from a 50-50 blend of viscosity-matched and density-matched mixture and incorporated the influence of coarsening on interfacial area and orientation on Q and $\mathbf{q}$ via scaling laws. Subsequent models retain the coarse-grained approach of the Doi-Ohta model, but modify how volume fraction, formation, coalescence, and pinching of drops and coarsening are included and thereafter influence interfacial area and orientation during flow,\cite{doi1991dynamics, minale2010models, larson1999structure}. A review by Minale \cite{minale2010models} provides a concise introduction to the different models inspired by Doi-Ohta formalism.

\textbf{Ellipsoidal drops based Maffettone-Minale model and its descendants} The phenomenological Maffettone and Minale model \cite{maffettone1998equation} that assumes that drops are always ellipsoidal and incompressible (drop volume is preserved). The MM model provides the evolution equation for a symmetric, positive-definite, second rank tensor $\mathbf{S}$ with with eigenvalues representing the square semiaxes of the ellipsoid. The tensor $\mathbf{S}$ measures deviations of the droplet from the spherical shape. The non-dimensional version of the equation obtained for scaled shape tensor $\bar{\mathbf{S}}=\mathbf{S}/{a}^2$ is as follows.
\begin{eqnarray}
    \label{eq MM model}
     \frac{d \bar{\mathbf{S}}}{d \bar{t}}-Ca (\bar{\mathbf{W}}\cdot \bar{\mathbf{S}}-  \bar{\mathbf{S}}\cdot \bar{\mathbf{W}}) 
     =  -{f_1}^{MM} [\bar{\mathbf{S}}-g(\bar{\mathbf{S}})\mathbf{I}] \nonumber&\\ +{f_2}^{MM}Ca (\bar{\mathbf{E}}\cdot\bar{\mathbf{S}}+ 
     \bar{\mathbf{S}}\cdot \bar{\mathbf{E}}) \,.
\end{eqnarray}
The left hand side of the MM equation involve Jaumann derivative rotating with vorticity and the right hand side has the influence of interfacial tension that drives recovery to the unperturbed shape (first term) and viscous drag that deforms the drops (second term). The MM model captures experimental results for simple shear and uniaxial and planar extensional flows as long as drops remain ellipsoidal (limited in Ca range). At O$({Ca})$, the MM model results recover Taylor's results for the following choices of the two coefficients:
\begin{eqnarray}
    \label{eq MM model coeff}
     {f_1}^{MM} = c_1(\lambda),     
     {f_2}^{MM} = c_0(\lambda)\,.
\end{eqnarray}
Minale's review \cite{minale2010models} lists and recaps a family of models that were developed inspired by the MM model. The descendant MM models incorporate the influence of coalescence and breakup to capture the response to high deformation rates with some success, and a few variants account for cases when one of the phases is non-Newtonian \cite{larson1999structure, minale2010models, maffettone1998equation, tucker2002microstructural}.  Models from MM family that use ellipsoid drops and allow for non-Newtonian drop or matrix phase as well as a few based on small deformation theory are able to capture the linear viscoelastic response of emulsions with reasonable success, but analysis and characterization of nonlinear viscoelasticity and response at high deformation rates remains challenging \cite{tucker2002microstructural, palierne1990linear, larson1999structure, minale2010models, cardinaels2016morphology}.

\subsection{Numerical methods for concentrated emulsions}
In this section, we enumerate representative numerical works on modeling semi-dilute to concentrated emulsion flows. We focus on the flow-induced microstructue of deformable drops in unbounded flows. Beyond the dilute regime, pairwise droplet interactions are affected by finite deformation of the drop interface allowing for hydrodynamic diffusion.  Droplet deformation in the near contact is the stabilizing mechanism against coalescence in the absence of van der Waals attraction \cite{loewenberg1997collision}. Scaling analysis for the near-contact motion between two clean drops within the lubrication regime shows slow algebraic film drainage $h/h_0 \sim \lambda/(\dot{\gamma} t)$ for $\dot{\gamma}t=O(1)$, where $h$ is the gap between the drops and $h_0$ is a reference, initial gap width. At long times, the internal circulation immobilizes near-contact motion preventing coalescence \cite{loewenberg1998conc}.  

As the volume fraction of the disperse phase or $\phi$ increases, as illustrated in Fig.~\ref{FIG:1}, many drop interactions become important and analytical treatment is limited. In this regime, detailed numerical simulations are often used to investigate flow-induced structuring and rheology of concentrated emulsions. The choice of numerical method depends largely on the system parameters (e.g., drop relaxation time, size distribution, and dispersed-phase concentration) and imposed flow conditions. Depending on the type of problem under investigation, for example, whether changes in drop topology or the near contact approach of droplet pairs are of interest, a balance among accuracy, resolution, meshing techniques, and computational cost plays a key role in selecting the appropriate numerical method. Complex fluid flows are inherently multiphysics problems governed by phenomena across lengthscales (e.g., from atomistic to continuum descriptions). Continuum numerical approaches for multiphase flows are typically divided in two main categories: interface capturing and interface tracking methods \cite{lowengrub2004MultiphaseMethods,prosperetti2009computational}. 

\textbf{Interface capturing and tracking methods}
Interface tracking methods explicitly track marker points on a grid or a mesh that fits the particle interface; classical examples are Boundary Integral Method (BIM) \cite{pozrikidis1992boundary} and Immersed Boundary Method (IBM) \cite{peskin2002immersed}. Alternatively, interface capturing methods (e.g., Volume of Fluid Method (VoF) \cite{hirt1981VoF}, Phase Field Method (PFM) \cite{cahn1958PhaseField}, and Level Set Method (LSM) \cite{osher2004levelset}) evolve a field variable across the computation domain where the interface is captured implicitly by a specific value of a field variable, for example, the contour of zeroes of the level set function \cite{gibou2018review}. At continuum scales, where volume-averaged material properties of the fluid are uniform, the interface between two immiscible fluids is often assumed to have zero thickness hence the definition of sharp or dividing interfaces \cite{basaran2023sharp}. Interface tracking methods are efficient and accurate in modeling sharp interfaces and are usually the method of choice when physical parameters vary strongly across an interface. However, topological changes (e.g., coalescence and breakup) are challenging and require highly detailed meshing schemes. Interface capturing methods handle topological changes naturally, whereas, interface tracking methods require additional numerical effort. For example, a numerical scheme called Arbitrary Lagrangian-Eulerian (ALE) has been proposed to track moving interfaces and resolve locally singular flows such as the pinching dynamics of fluid droplets  \cite{basaran2023sharp}. Usually, ALE approach is combined to a spatial and time discretization schemes (e.g., Garlerkin Finite Element (GFEM) and Finite Differences (FDM)) and to specialized meshing techniques to resolve the rapid variations of the pressure and axial velocity in the neck region during pinching \cite{kistler1984coating,basaran2002smallcale,basaran2023sharp}. The challenge of using interface capturing methods to model physical systems where material properties are discontinuous across an interface, may be overcome by a hybrid approach of interface capturing methods and immersed interface methods or ghost fluid methods \cite{tryggvason2011DNS,hu2015hybrid,salac2023ModelingSlipDrops}. 

\textbf{Particle-based models} At mesoscopic length scales bridging the gap between molecular dynamics and continuum simulations, coarse-grained particle-based models (e.g., Dissipative Particle Dynamics \cite{hoogerbrugge1992DPD}) or kinetic-based models (e.g., Lattice Boltzmann Method \cite{kruger2017lattice}) are usually employed giving access to additional physics compared to continuum-based approaches such as the Boundary Integral Method or Level Set Method. However, both mesoscopic methods require large computational costs to achieve refined grid resolution typically needed in handling near-contact interactions among suspended particles accurately.

\begin{figure*}[h!]
	\centering
	\includegraphics[scale=.51]{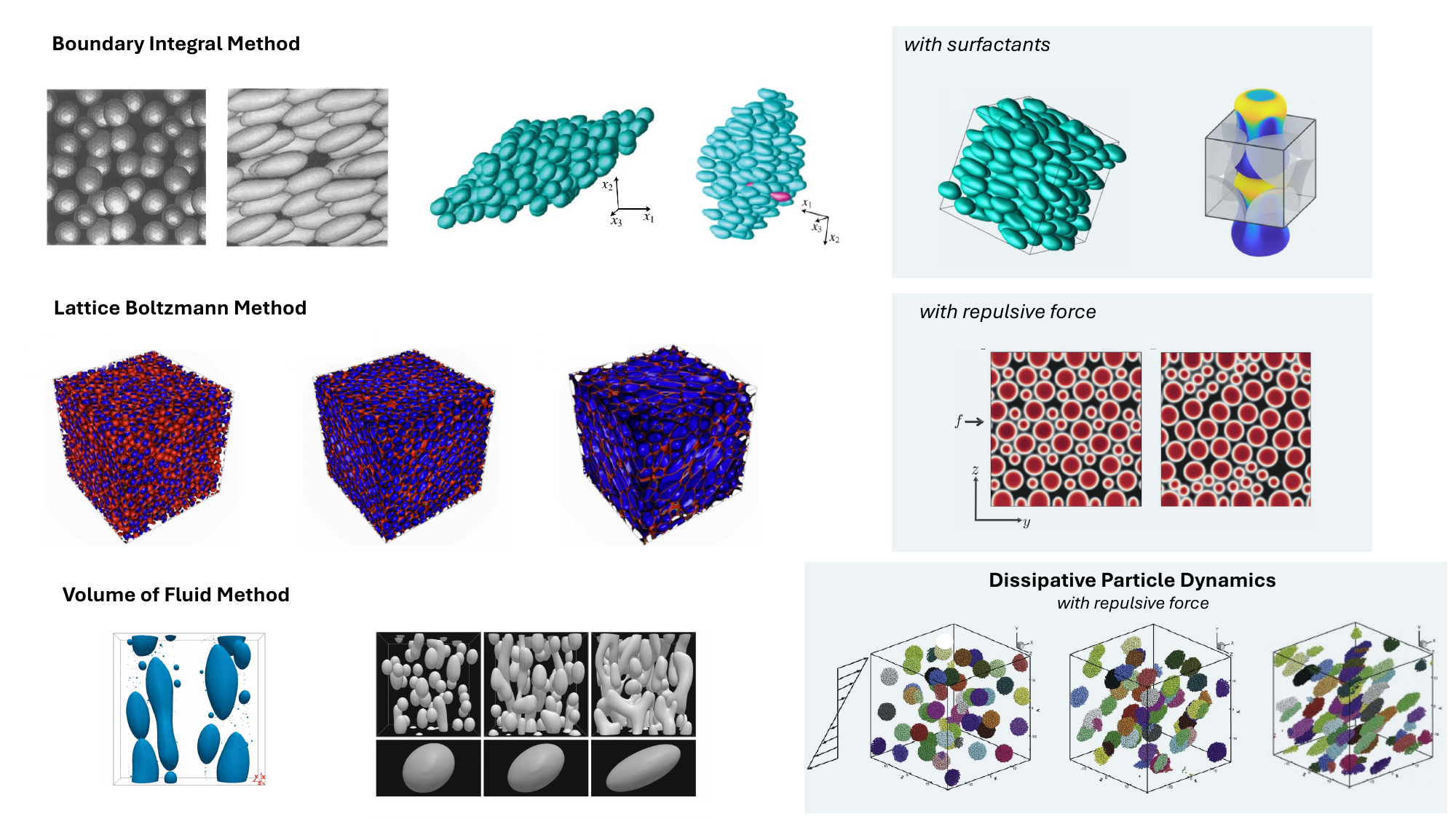}
        \put(-488,248){$(a)$}
        \put(-340,242){$(b)$}
        \put(-180,242){$(c)$}
        \put(-100,242){$(d)$}
        \put(-488,152){$(e)$}
        \put(-180,152){$(f)$}
        \put(-475,55){$(g)$}
        \put(-384,55){$(h)$}
        \put(-215,65){$(i)$}
	\caption{Summary of representative numerical works on concentrated, highly concentrated, and dense emulsions. The images are adapted from references using Boundary Integral Method for small-scale \cite{loewenberg1996numerical} (a) and large-scale simulations \cite{zinchenko2002shear} (b) of clean drops in shear flow; emulsion flow of surfactant-covered droplets in shear flows \cite{zinchenko2017general} (c) and  through structured domains \cite{zinchenko2022surfacPorous} (d), Lattice Boltzmann Method for flowing emulsions where stabilization against coalescence can be tuned by a repulsive force \cite{girotto2024lagrangian} (e), Lattice Boltzmann Method for jammed, dense emulsions of slightly deformed droplets \cite{negro2023yield} (f), Volume of Fluid simulations of flowing concentrated emulsions accounting for irreversible topological transitions \cite{Brandt2020numericalshearbands} (g) and \cite{rosti2019numerical} (h); and Dissipative Particle Dynamics for concentrated emulsions of droplets in shear flow \cite{pan2014dissipative} (i). Details of each method can be found in Fig.~\ref{FIG:Numerical Methods}.}
	\label{FIG:4}
\end{figure*}

\begin{figure*}[h!]
	\centering
		\includegraphics[scale=.52]{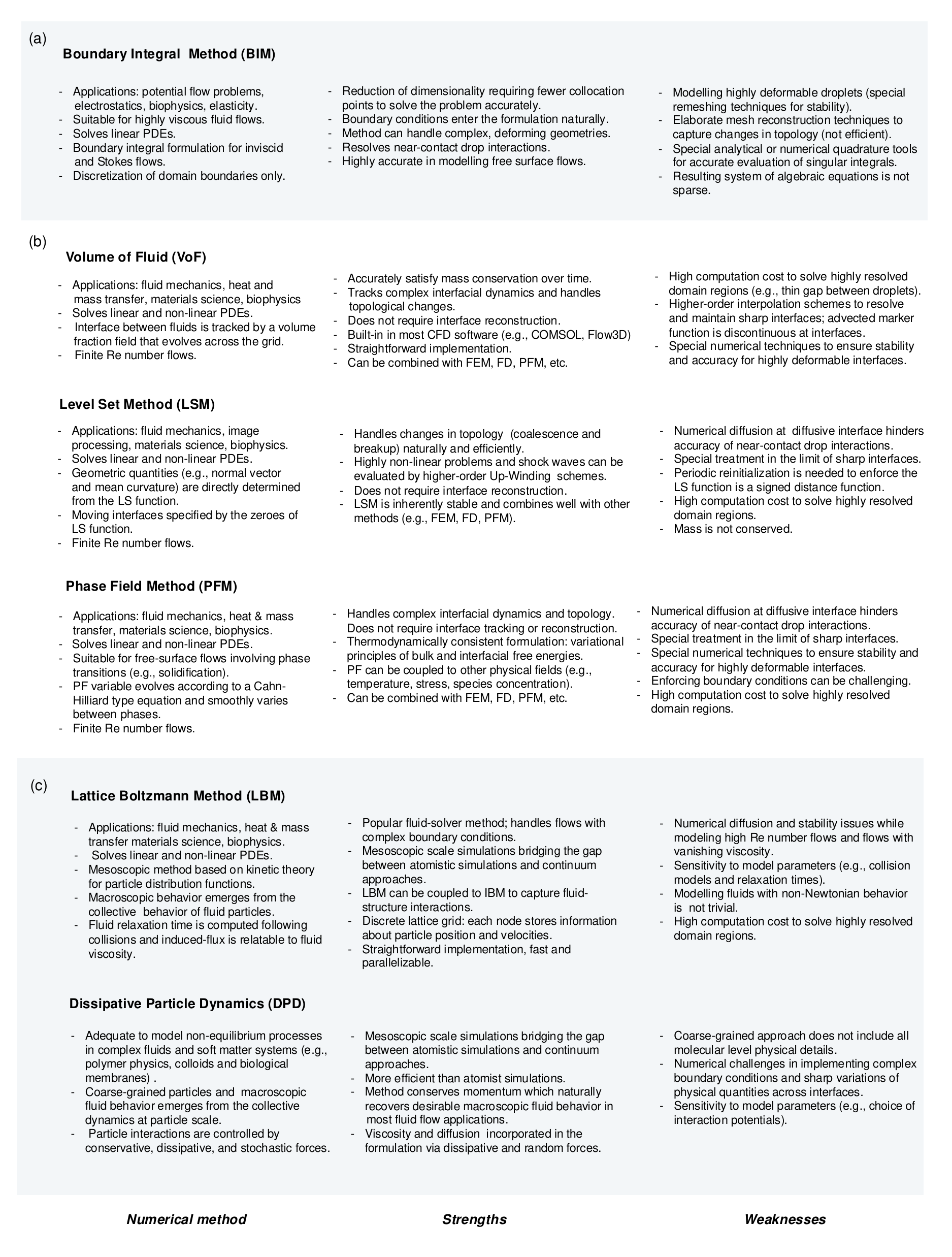}
	\caption{Mapping of representative numerical methods typically used in simulations of concentrated emulsion flows. The three areas (a), (b), and (c) refer to interface tracking, interface capturing, and particle based methods, respectively. First column shows a general description of each method. The last two columns highlight strengths and weaknesses. Abbreviations used: Finite Element Method (FEM), Finite Difference (FD), Immersed Boundary Method (IBM), Computational Fluid Dynamics (CFD), Partial Differencial Equations (PDEs), Reynolds number (Re), and numerical methods as indicated. }
	\label{FIG:Numerical Methods}
\end{figure*}

Figure \ref{FIG:4} highlights representative numerical results of concentrated to dense emulsions using some of the methods listed in Fig.~\ref{FIG:Numerical Methods}. Figure \ref{FIG:Numerical Methods} compiles a descriptive map of representative interface tracking, interface capturing, and coarse-grained mesoscopic numerical approaches used in modeling of drop deformation in multiphase flows. The first column under each method lists the range of applicability and main features of each method, the second and third columns summarizes their strengths and weaknesses, respectively. Simulating drop deformation in flowing emulsions depend highly on system parameters and on the questions under investigation. For example, if one is interested in evolving the microstructure of an emulsion at low-Reynolds number conditions where topological changes are not relevant to the system dynamics, then using BIM would be a method of choice in terms of accuracy and efficiency compared to interface capturing methods such as VoF or LSM or even coarse-grained methods like LBM. Alternatively, VoF or LSM are more efficient and stable methods compared to BIM to resolve creeping flows of concentrated emulsions where changes in topology are present since no remeshing is needed to handle coalescence or breakup events. However, handling near-contact regions between droplets hinders accuracy and increases the computational cost when using LSM and VoF. Here, our goal is to highlight that each emulsion flow problem introduces inherent physical challenges and should be probed by the appropriate numerical tool. The summary of methods listed in Fig. \ref{FIG:Numerical Methods} could be used as a guide.

For a comprehensive review on numerical methods used in modeling interfacial rheology including viscoelastic effects at the interface and sharp-interface methods to solve free-surface flows, the reader is directed to Refs. \cite{jaensson2021computational} and \cite{basaran2023sharp}, respectively; and to Ref. \cite{prosperetti2009computational} for more details on other computational methods for multiphase flows.

\textbf{Examples of numerical works on concentrated emulsions} Loewenberg \& Hinch \cite{loewenberg1996numerical} used boundary integral simulations and presented one of the first attempts to simulate small-scale numerical analysis of concentrated emulsion flows of clean, deformable drops with dispersed-phase volume fraction $\phi\leq 30\%$. The results showed a strong shear-thinning behavior, with large positive first and negative second normal stress differences, where typically $\vert N_1 \vert>\vert N_2 \vert$. This rheological response is illustrated by the microstrucuture anisotropy shown in Fig.~\ref{FIG:4}(a) where droplets are more deformed and aligned with the flow direction (left image), whereas in the vorticity direction the drops are closely packed (right image). Elongation of the droplets in the flow direction promotes large $N_1$ and facilitates the motion of drops past each other. This droplet arrangement reduces the collisional cross-section and local viscous dissipation leading to a shear thinning behavior. A similar system of interacting droplets in concentrated emulsions with $\phi<30\%$ has been investigated including inertial effects on the emulsion rheology and flow-induced drop structure \cite{pan2014dissipative}. The authors used Dissipative Particle Dynamics method where droplets are stabilized against coalescence by a strong repulsive force as illustrated in Fig.~\ref{FIG:4}(i); breakup events are not considered.

More recent studies address flow-induced structuring and rheology of highly-concentrated emulsions below critical jamming conditions \cite{zinchenko2002shear,zinchenko2015extensional}. Zinchenko \& Davis \cite{zinchenko2015extensional} used a large-scale boundary integral simulation to probe the 
rheology of highly-concentrated emulsions in flows with nontrivial kinematics. Large strains were assumed and disperse-phase 
volume fraction varied in the range $0.45<\phi<0.55$. The simulations used 400 drops per periodic cell and improved upon earlier works from the same group \cite{zinchenko2000efficient,zinchenko2002shear}. A snapshot of a periodic cell is shown in \ref{FIG:4}(b). The authors propose a five-parameter, generalized Oldroyd model 
where the variable parameters are determined from viscometric and extensionmetric base flows. For example, shear viscosity, first- and second-normal stress differences are calculated from shear flows, and extensional viscosity and stress cross-difference from extensional flows. Long-time averaged material properties in mixed shear and pure extensional flows retain the qualitative features obtained in small-scale simulations of monodisperse emulsions $\phi\leq30\%$ \cite{loewenberg1996numerical}. 


Numerical analysis of drop-scale deformation and bulk rheology beyond the class of clean, deformable droplets have been mostly restricted to dilute to semi-dilute regimes accounting for surfactant-covered drops or drops with surface viscous dissipation \cite{li1997effect,sorgentone2018highly,pimenta2021study}. Recently, Zinchenko \& Davis \cite{zinchenko2017general} extended their numerical scheme for highly concentrated emulsion of clean drops \cite{zinchenko2015extensional} to drops covered with insoluble surfactants \cite{zinchenko2017general} in shear and extensional flows. They studied emulsion flows with dispersed-phase volume fractions $0.45<\phi<0.6$, viscosity ratio $0.25<\lambda<3$, and surfactant elasticity $0.05<\beta<0.2$. Sophisticated meshing schemes needed to capture highly deformed droplets in nearly jammed dense emulsions and numerical resolution of the near contact phenomena of approaching droplets are challenges faced by researchers in this field. A representative snapshot of highly-concentration emulsion of surfactant-covered droplets in shear flow is shown in Fig.~\ref{FIG:4}(c). Figure \ref{FIG:4}(d) shows BIM simulations a pair of highly-deformable surfactant-covered droplets flowing through a pore geometry; the color gradient along the surface indicates regions of different surfactant concentration.

\textbf{Influence of drop coalescence and breakup} Transient evolution of the emulsion micro-structure in concentrated emulsions including changes in droplet topology (e.g., breakup and coalescence events) remains an open area of research. The critical effect of flow-induced droplet breakup and fragmentation on the microstructure and rheology of emulsions \cite{cristini2003drop,van2008review}, including wall-effects \cite{cristini2004review,bazhlekov2006numerical,janssen2010generalized,guido2011shear,kapiamba2022mini,roure2023numerical}, external force fields \cite{miksis1981shape,cunha2018field,raysa2024}, non-Newtonian contributions from either the dispersed or continuous phases (e.g., viscoelastic, power-law, elastoviscoplastic) have been reviewed or studied elsewhere  \cite{guido2011shear,rosti2021shear,wang2023deformation,zhang2024numerical,raysa2024}.

Coalescence and breakup events may coexist in confined emulsion flows leading to nontrivial rheology. For example, shear bands which are regions of high and low droplet concentrations in the vorticity and flow direction, respectively, have been observed in numerical experiments \cite{caserta2012vorticity,Brandt2020numericalshearbands}. Figure \ref{FIG:4}(g) shows a snapshot of the droplet microstructure in VoF simulations adapted from Ref.~\cite{Brandt2020numericalshearbands}. Rosti et al. \cite{rosti2019numerical} determined the effective viscosity of concentrated emulsions using a 3-dimensional VoF method for volume fractions in the range $10^{-3}<\phi<0.3$ and capillary number $0.1<Ca<0.3$. Coalescence events lead to a non-monotonic variation of effective viscosity with $\phi$, with a peak around $\phi \approx 0.20$. The representative droplet shape distribution observed in their VoF simulation is shown in Fig.~\ref{FIG:4}(h). Recently, Girotto et al. \cite{girotto2024lagrangian} used mesoscopic Lattice-Boltzmann method to study the evolution of the microstructure of emulsions as the disperse-phase volume fraction increases from semi-dilute to jammed configurations. The authors included coalescence and breakup events and further studied aging dynamics effects after the flow is stopped. An evolution of the emulsion droplet network as the concentration increases is shown in Fig.~\ref{FIG:4}(e). For a comprehensive review on numerical aspects and recent progress on the modeling of deformable particles in flows using the Lattice-Boltzmann method see Ref.~\cite{silva2024lattice}. Peterson et al. \cite{peterson2023viscoelastic} proposed a generalized framework model for droplet breakup in dense emulsion flows using a population balance model coupled to droplet shape evolution.

\section{Jammed dense emulsions with polygonal drops in a network of films} \label{sec: dense}

On increasing the volume of the dispersed-phase or $\phi$ beyond the highly concentrated regime of flowing emulsions discussed in section \ref{sec: concentrated}, the rheological response shows manifestation of a yield stress, implying flow occurs only after minimum threshold value of stress (or applied force) is exceeded. The magnitude of yield stress and beyond yield stress, the flow behavior shows high sensitivity to the positional structure, size, shape, interparticle forces, and polydispersity of droplets. In this regime, an emulsion of repulsive droplets (stabilized against coalescence) transitions from amorphous glass-like behavior for $\phi_g\approx 0.58$ to a jammed, dense regime at $\phi\approx\phi_{RCP}$ where the microstructure is dense and randomly packed and $\phi_{RCP}\approx 0.64$. In the limit as $\phi\to 1$, the drops get compressed into polygonal shapes. The deformed drops are separated by thin films of the continuous phase fluid, and the films that intersect at Plateau borders thus develop a microstructure or a castle of polyhedral shapes characteristic of dry foams \cite{mason1999rheologyCOCIS,larson1999structure,KimMason017advances, langevin2020emulsions, foudazi2015physical, larson1997elastic}. In this section, we focus on the structure and rheology of jammed dense emulsions where the droplets are densely packed showing a solid-like behavior under weak loading, and a fluid-like behavior beyond an effective yield stress \cite{princen1983rheologyI,princen1989rheologyIV}. 

Dilute and nondilute flowing emulsions, as discussed in sections \ref{sec: dilute}-\ref{sec: concentrated}, exhibit a non-Newtonian rheology and viscoelastic response, and their elasticity is attributed at the drop level to the interfacial tension dependent shape relaxation time. Jammed dense emulsions show a viscoplastic response to imposed bulk stresses, such that flow only occurs after yield stress is exceeded. The empiricial Herschel-Bulkley model is often used for capturing the flow behavior for a complex fluid that displays a yield stress and flows with a power law relationship between stress and deformation rate above yield stress. The three parameter HB model includes a power law exponent, $n$, consistency, $K$, and yield stress, $\tau_{Y}$,  and can be written as
\begin{equation}
    \label{eq Herschel Bulkley}
    \tau=\tau_{Y}+K{\dot{\gamma}^n}=\tau_{Y}+ \tau_{v}{(\dot{\gamma})}\,
\end{equation}
The HB model is a generalization of the Bingham model that includes only two parameters (as the power law exponent equals 1). More elaborate models such as SGR (soft glassy rheology) model, mode coupling theory can be derived to capture the rheological behavior of yield stress materials by accounting for local traps or local rearrangement zones in dispersions containing densely packed drops or particles, as detailed elsewhere \cite{bonn2017yield, larson1999structure, cates2004tensorial}.

The viscoplastic behavior may be qualitatively defined using the Bingham number, $Bn$ 
\begin{equation}
    \label{eq Bingham number}
    Bn=\frac{\tau_{Y}}{\tau_c}\,,
\end{equation} 
which is simply the ratio of yield stress $\tau_Y$ and an imposed characteristic stress, $\tau_c=\mu U/L$. Here $\mu$ is characteristic viscosity, $U$ and $L$ are characteristic velocity and length scale, respectively. 

Under small strains compared to $\tau_Y$, dense emulsions show a jammed, solid-like behavior with elastic modulus given by 
\begin{equation}
    \label{eq elatic modulus}
    G\approx \frac{\sigma}{a_{32}} \phi^{1/3}(\phi-\phi_0) \,,
\end{equation}
where $\sigma$ is interfacial tension coefficient, $a_{32}=3V/A$ is an volume-to-surface-area mean drop radius, and $\phi_0\approx 0.71$ is the limiting volume fraction at which the percolation of the droplet network collapses. The rheology of dense emulsions of non-coalescing droplets including typical flow curves and characteristic viscoelastic behavior described by the storage, $G'$, and loss moduli, $G''$, subject to linear and non-linear viscoelastic flowing regimes and has been well documented in reviews and papers\cite{mason1999rheologyCOCIS,foudazi2015physical,KimMason017advances,kapiamba2022mini, langevin2020emulsions, datta2011rheology}, where most of the works are experimental. Theory and numerical aspects of the problem remain an active area of research.

The measurement or observation of an apparent yield stress in jammed dense emulsions and suspension of particles with a relatively wide range of interaction is much easier than describing the underlying mechanism involving dynamics of dispersed drops in the case of emulsions.\cite{vlassopoulos2014tunable, nelson2019designing, bonn2017yield, geffrault2021extensional, nicolas2018deformation, clara2015affine, cao2021rheology} The collapse of the amorphous glass-like microstructure signals the transition to a fluid-like behavior where a classical empirical model by Princen and Kiss \cite{princen1989rheologyIV} for the yield stress is
\begin{equation}
    \label{eq Princen Yield Stress}
    \tau_{Y}=\frac{\sigma}{a_{32}} \phi^{1/3} Y(\phi) \,,
\end{equation}
and $Y(\phi)$ is an empirical relation showing a logarithmic dependence on $\phi$ \cite{princen1989rheologyIV}. Several models are proposed as detailed in the review by Kim and Mason \cite{KimMason017advances}. Figure 6 illustrates that two empirical models capture the trends observed experimentally for $\phi$ dependent increase in modulus and yield stress. Details including the properties of dispersed and suspending fluid, the expression for computing the two quantities and values used for different constants are listed in the Appendix for completeness. For emulsions that display yield stress, recent experiments using gravity-based rheometry show the possibility of measuring both an extensional yield stress and the power law relation between extensional stress and strain rate using analysis of dripping, though challenges remain in quantitatively describing the underlying mechanisms for strong flows where droplet deformability probably plays a role.\cite{kibbelaar2023towards, geffrault2021extensional, niedzwiedz2010extensional, nikolova2023rheology}

\begin{figure*}[h!]
	\centering		\includegraphics[scale=.41]{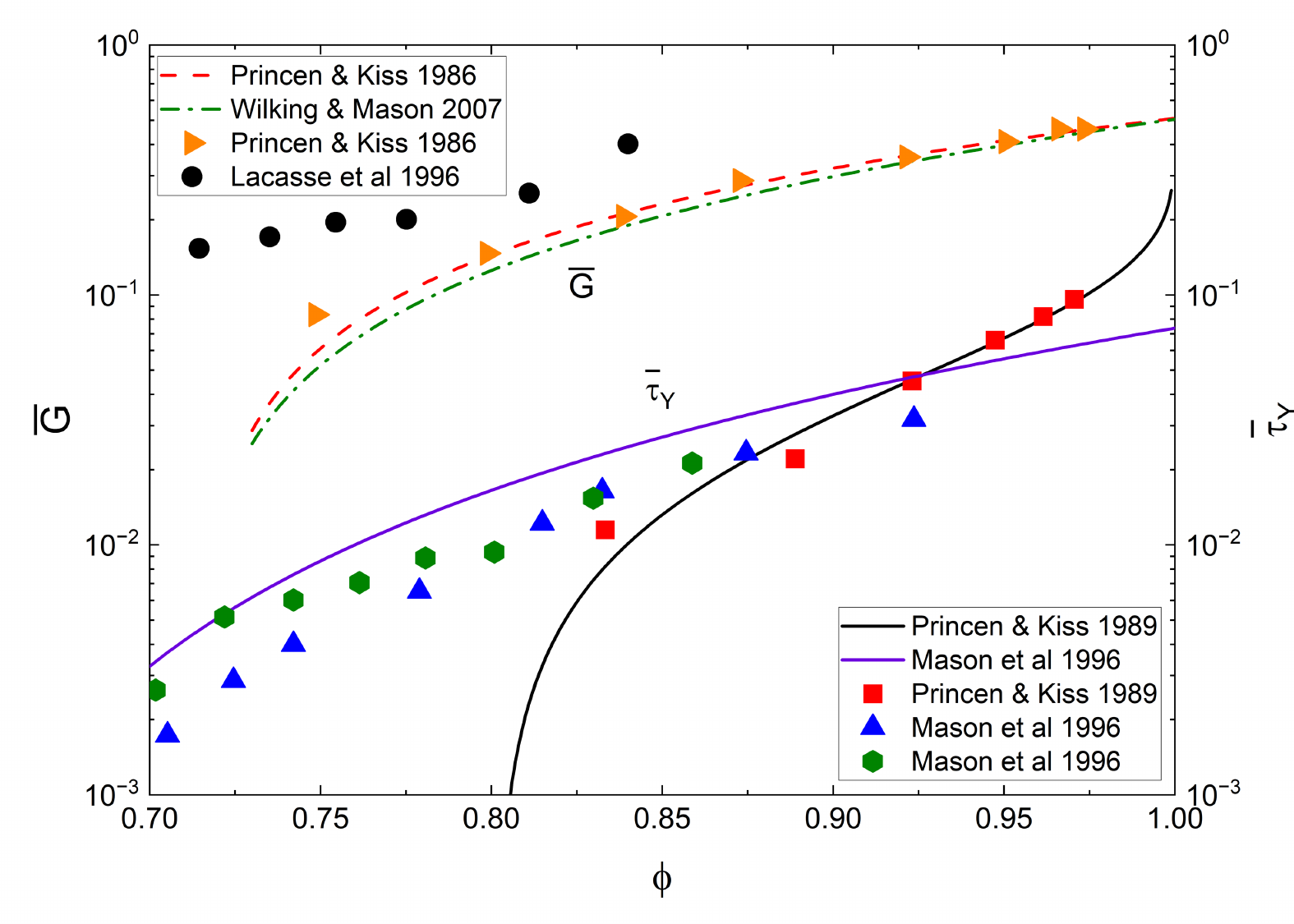}
 \caption{Comparison between elastic modulus and yield stress empirical models for jammed dense emulsions. Data sets obtained from Refs.~\cite{princen1986rheology} and  \cite{lacasse1996model} for elastic modulus and from Refs.~\cite{princen1989rheologyIV} and \cite{mason1996yielding} for yield stress. Empirical models for elastic modulus obtained from Refs.~\cite{princen1986rheology} and \cite{wilking2007irreversible}, and for yield stress from \cite{mason1996yielding} and \cite{princen1989rheologyIV}. Both elastic modulus and yield stress are normalized by a characteristic capillary stress $\sigma/a$. See Appendix \ref{Appendix: dataset fig 3} for details on the datasets and models used.}
	\label{FIG:6}
\end{figure*}

Denkov and coworkers \cite{denkov2008viscous} argued that the second term or the viscous stress contribution, $\tau_{v}(\dot{\gamma})$ for yielded emulsions can be attributed to the energy dissipation in thin films between neighboring drops sliding along each other. Their model anticipates a power law exponent $n$ = 1/2 if disjoining pressure is neglected, and explains why viscous stress and shear viscosity exhibit $Ca^{1/2}$ and $Ca^{-1/2}$ scaling, respectively for flowing emulsions. An extended version of the model suggests $n$ < 1/2 if interfacial dissipation plays a role and $n$ > 1/2 if disjoining pressure exerts an influence. The model appears to capture the diversity in power law exponents observed experimentally  in flowing emulsions \cite{tcholakova2008theoretical, denkov2008viscous}.

\textbf{Numerical studies of jammed dense emulsions} Emulsions display $\phi$ dependent yield stress, and is often used by experimentalists as a model system for investigating rheological response. Numerically modeling jammed dense emulsions proffers a similar opportunity with the advantage that changes in microstructure below and above yield stress in response to applied stress can be visualized and analyzed, as shown in a recent numerical investigation by Negro et al. \cite{negro2023yield}. The authors numerically investigated in 2D the yield stress and flow behavior of a model emulsion that contains an amorphous deformable non-coalescing droplets embedded in a Newtonian fluid, as summarized below.

Negro et al. \cite{negro2023yield} evolved the droplet dynamics using 2D hybrid Lattice-Boltzmann method and computed hydrodynamics by following the evolution of phase field variables and velocity of the suspending fluid using the Cahn-Hilliard equation. The droplets are stabilized against coalescence by a soft repulsion force providing for a weak overlap between droplets and forming a percolated microstructure. The  model system of densely packed droplets of conserved area initially lies in an amorphous, immobile glass-like state in response to an external forcing, $f$ or pressure difference in a parabolic flow. When the forcing is greater than a critical value $f_c$, the percolated network yields and the microstructure orders along the flow direction. Even for $f<f_c$, numerical results indicate the continuous fluid permeates the immobile droplet network and hence the effective viscosity is large but finite. Yielding transition is marked by droplet mean velocity fluctuations and stick-slip fluid motion. An analysis of bidisperse systems of small and large species reveals a similar phase transition occurs for $f>f_c$. In this regime, yielding is followed by an ordered microstructure where large species accumulated near the centerline of the pressure-driven flow and small species are marginated, as shown in Fig.~\ref{FIG:4}(f). This behavior is reminiscent of flow-induced structuring in the bulk and near the boundaries of dilute to concentrated suspensions given by a balance among hydrodynamic diffusion, deformation-induced drift velocity, and local velocity gradient fluxes 
\cite{LeightonAcrivos1987,LeightonAcrivos1987A,SmartLeighton1989,Phillips1992,Nott1994,daCunha1996,loewenberg1997collision,Narsimham2013,RiveraGraham2016,qi2017theory,reboucas2022pairwise}.

\section{Challenges, opportunities, and prognosis}\label{sec: challenges}
\sloppy
Over the past century, the progress in describing the physicochemical origins of the flow behavior of emulsions reflects progress in describing soft matter physics, thermodynamics, intermolecular and surface forces, interfacial properties, and drop deformation, breakup and coalescence. Despite progress, designing more sustainable, cost-efficient, or functional emulsion-based formulations remains challenging as many fundamental scientific problems arise. The macromolecular, supramolecular and particulate ingredients can alter the rheology of dispersed or suspending fluids and influence interfacial properties, affecting stability, application and processing of emulsions. The review captures some highlights from the current state-of-the-art in modeling shear rheology of emulsions containing Newtonian drops in Newtonian continuum phase with a Newtonian interface. Making any of the three non-Newtonian introduces conceptual, characterization and modeling challenges. Additional open questions are encountered in the following contexts, where we restrict discussion to theoretical and computational challenges only.

\begin{figure*}
	\centering
		\includegraphics[scale=.62,angle=90]{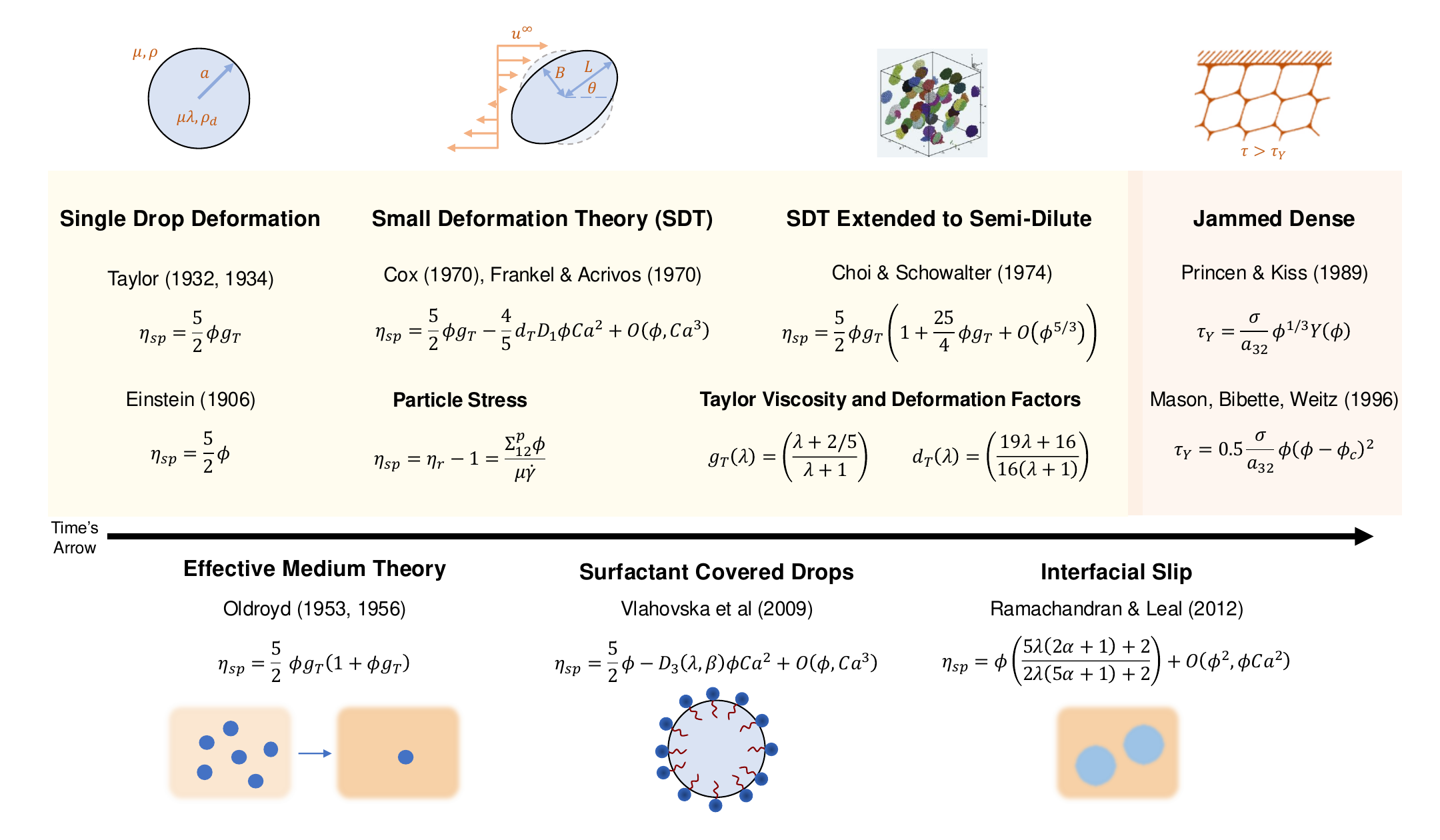}
	\caption{A compendium of analytical expressions for relative viscosity for dilute to nondilute emulsions derived using small deformation or effective medium theory and phenomenological relations for yield stress in jammed dense emulsions. The publication years are included for each model to highlight the milestones corresponding to time's arrow.}
	\label{FIG:viscosity models}
\end{figure*}

 \textbf{Extensional rheology response} requires a careful consideration of large changes in drop shapes, which enhances the possibility of breakup or coalescence of drops and microstructure changes, that in turn influences the response of streamwise velocity gradients\cite{delaby1995elongation}. For nondilute emulsions, there is also a pronounced lack of experimental data that can be used to benchmark theoretical methods. There is a lack of in-situ techniques that can be used to measure extensional viscosity while visualizing the evolution of drop shapes and microstructure in response to practically relevant deformation rates  \cite{nikolova2023rheology,niedzwiedz2010extensional,dinic2017pinch}.

\textbf{Influence of non-Newtonian interfacial rheology} Connecting the emulsion rheology response to the specific measures of interfacial rheology response in dilatational, shear, elastic, bending and torsion modes remains a challenge that can benefit from combination of modeling and experimental studies\cite{erni2011deformation}. Adsorbed layers of proteins, surfactants, polymers, particles, and lipids can display interfacial properties ranging from mobile to rigid, spanning theories discussed herein to describing drops with clean interfaces to elastic interfaces (like capsules) \cite{erni2011deformation, erni2011emulsion, langevin2020emulsions, barthes2011modeling, barthes2016motion, fischer2007emulsion, narsimhan2019shape}.

\textbf{Viscoelastic dispersed or suspending fluid} Despite significant progress in analyzing drop deformation for cases with either phase or both phases are non-Newtonian, the constitutive models and numerical studies described in this contribution are inadequate for capturing the rheological response at high deformation rates for emulsions containing viscoelastic suspending fluids or viscoelastic droplets in a Newtonian suspending medium. The two-way coupling of bulk elastic stresses to the interfacial stress jump across the interface can be highly nonlinear, introducing challenges in modeling multiphase flows containing moving boundaries. The effect of flow-induced cross-stream migration and deformation of droplets or capsules in viscoelastic background fluids on the rheology of dilute to concentrated suspensions also remains an open area of research \cite{langevin2022motion,NeoShaqfeh2024Capsules, langevin2020emulsions, guido2011shear}. 


\textbf{Bubbly fluids} Theoretical and experimental investigations on the transient rheology of bubble suspensions remain an active area of research \cite{ohie2024bubblerheology,mitrou2024linearbubbles}. In the limit of emulsions containing bubbles as the dispersed phase ($\lambda\to 0$), Rust \& Manga \cite{rust2002bubbleOrientationShape,rust2002viscosity} compared small and large deformation theories and numerical calculations to experimental results on the shape, deformation and effective viscosity of surfactant-free bubbly suspensions. Around the same time, Llewellin et al. \cite{llewellin2002rheology} proposed a constitutive model to study the transient rheology of polydisperse bubble suspensions derived using the small deformation theory of Frankel \& Acrivos \cite{frankel1970constitutive}  that reduces to a three-parameter Jeffrey-like model typically used to characterize the rheological response of viscoelastic liquids, as shown in Eq.~\refeq{eq constitutive Jeffreys form} \cite{barnes1994rheology}.

\textbf{Role of deformation and processing history, including emulsification} Changes in drop sizes and size distributions, and microstructure have a direct impact on the overall flow properties of emulsions. Modeling different emulsification methods \cite{haakansson2019emulsion, haakansson2023emulsifier, vankova2007Part1, vankova2007Part2, khakhar1986deformation} and polydisperse droplets to emulate the formulations used in personal care, food, or industry grade emulsions requires a deeper dive into the rheology and thermodynamics of multicomponent systems \cite{windhab2005emulsion, dardelle2014three, fick2023interfacial, langevin2020emulsions, erni2011emulsion, bibette2003emulsion, tadros2016emulsions, mcclements2017recent}.

\textbf{Yielding and microstructural evolution of jammed dense emulsions} Further research is needed to elucidate the effects of changes in local drop size, shape and number density, and topological changes involving interconnected thin films on the bulk rheology of emulsions, especially if disjoining pressure, and interfacial rheology effects are to be included, and if non-Newtonian phases are involved. \cite{geffrault2021extensional, kibbelaar2023towards, nikolova2023rheology, bonn2017yield, larson1999structure, chatzigiannakis2021thin, cao2021rheology, larson1997elastic, KimMason017advances}.

We close this overview with Figure \ref{FIG:viscosity models} that highlights the key equations and the underlying physicochemical considerations described in this contribution. Figure \ref{FIG:viscosity models} includes a timeline of analytical relative viscosity models from dilute to semi-dilute systems and phenomenological relations for yield stress of jammed dense emulsions. We offer this survey of theoretical and numerical modeling of emulsions rheology to the scientific community, with the awareness that despite this remarkable progress, many practical problems remain in producing, storing, processing, and designing emulsions. We anticipate that advances in numerical and computational methods, and the emergence of exciting problems and consumer/industry-driven quests involving food and personal care emulsions made with sustainable ingredients will drive the field in the near future.



\section*{Appendix}
\appendix
\section{Small deformation theory: clean drops distortion tensor} \label{appendix: small deformation theory}
In the limit when suspended neutrally bouyant, clean droplet deviates from sphericity only slightly, the droplet surface is given by \cite{rallison1980note}
\begin{equation}
    \label{eq drop surface 3}
    S(t)=r(t)-a \left(1+ \epsilon \frac{ \bx \cdot \mathbf{A}(t) \cdot \bx }{r^2}\right) + O(\epsilon^2) = 0
\end{equation}
where $\epsilon \ll 1$ is a small parameter, $\mathbf{A}$ is the shape distortion tensor, $a$ is the radius of the undeformed, spherical droplet, and $r=(\bx \cdot \bx )^{1/2}$ is the radial position measured from the droplet center. The shape distortion tensor is a measure of elongations of material lines of a deformable body,
\begin{equation}
    \label{eq elongation}
    \frac{ds}{ds_0} = \left(1-2  \epsilon \frac{\mathbf{x}}{r} \cdot \mathbf{A} \cdot \frac{\mathbf{x}}{r}\right)^{-1/2}\,,
\end{equation}
where $ds_0$ and $ds$ are the length of a material line at a reference and current configuration. By definition, the shape distortion tensor $\mathbf{A}=\left(\mathbf{I}-\mathbf{B}^{-1}\right)/2$, where $\mathbf{B}=\mathbf{F} \cdot \mathbf{F}^T $ is the left Cauchy-Green deformation tensor and $\mathbf{F}$ is the deformation gradient tensor. The deformation gradient tensor is a second order tensor that maps the deformation of material lines from a reference to a current configuration $d\mathbf{x}=\mathbf{F} \cdot d\mathbf{x}_0$ whose components are $F_{ij}=\partial x_i/\partial x_{0j}$ \cite{spencer2004continuum}. Taking a Taylor series expansion of the right-hand-side of Eq.~\refeq{eq elongation} yields,
\begin{equation}
    \label{eq elongation expanded}
    \frac{ds}{ds_0} = 1+ \epsilon \frac{\mathbf{x}}{r} \cdot \mathbf{A} \cdot \frac{\mathbf{x}}{r}+\epsilon^2 \frac{3}{2}\frac{\mathbf{x}\mathbf{x}:\mathbf{A} \mathbf{A}:\mathbf{x}\mathbf{x}}{r^4}+\ldots\,.
\end{equation}
Equation \refeq{eq drop surface 3} is obtained from relation \refeq{eq elongation expanded} noting that perturbations in the droplet shape are captured by $r(t)=a (ds/ds_0)$ and hence the surface is defined by $S(t)=r(t)-ds/ds_0\equiv 0$ to leading order in $\epsilon$.

The solutions to Eqs.~\refeq{eq Stokes continuous}-\refeq{eq Stokes drop} are obtained assuming a spherical shape by, for example, superposition of vector spherical harmonics \cite{leal2007advanced}. To leading order, shape distortions are captured in the definition of the normal vector $\mathbf{n}=\nabla S(t)/\vert \nabla S(t)\vert$ such that \cite{oliveira2015emulsion}
\begin{equation}
    \label{eq normal vector}
    \mathbf{n} = \frac{\mathbf{x}}{r} -2 a \, \epsilon \left[\frac{\mathbf{A} \cdot \bx}{r^2}-\frac{\bx\left(\bx \cdot \mathbf{A} \cdot \bx\right)}{r^4} + O(\epsilon^2)\right] \,,
\end{equation}
and hence appears in the calculation of the mean curvature, $H$, given by Eq.~\refeq{eq mean curvature}. Enforcing boundary conditions \refeq{eq BC velocity jump} and \refeq{eq BC traction jump} at the droplet interface, the leading order interfacial velocity reduces to 
\begin{equation}
    \label{eq interfacial velocity}
    \mathbf{u}_s=\mathbf{W}\cdot \bx +c_0(\lambda) \bE \cdot \bx - \frac{\sigma}{\mu a} c_{1}(\lambda) \,\epsilon \mathbf{A} \cdot \bx \,,
\end{equation}
where $c_0(\lambda)=5/(2 \lambda +3)$, $c_1(\lambda)=40(\lambda +1)/[(19 \lambda +16)(2 \lambda+3)]$, and $\mathbf{E}$ and $\mathbf{W}$ are the imposed-flow rate-of-strain and vorticity tensors, respectively, i.e., $\mathbf{u}^{\infty}=(\mathbf{E}+\mathbf{W})\cdot \bx$. Inserting Eq.~\refeq{eq drop surface} into the kinematic boundary condition \refeq{eq BC kinematic} written in the form $DS(t)/Dt=0$, where $D/Dt=\partial/\partial t + \mathbf{u} \cdot \nabla$ is the material derivative \cite{batchelor1967introduction}, and using the approximation that $D(\mathbf{x}/r)/Dt\approx \mathbf{W}\cdot \mathbf{x}/r$, $D r/Dt =(\mathbf{x}/r)\cdot \mathbf{u}_s$, and that $\mathbf{W}$ is anti-symmetric yields the evolution equation for the distortion tensor \cite{rallison1984deformation,oliveira2015emulsion}.

\subsection{Second-order deformation theory coefficients: clean droplets} \label{appendix: small deformation coeffs}
For completeness, the coefficients appearing in Eqs. \\ \refeq{eq shear stress 12 clean}-\refeq{eq N2 emulsion} for clean drops are listed below \cite{vlahovska2009small},
\begin{eqnarray}
    \label{eq rheology coeff D0 clean}
    D_0(\lambda)=\frac{(19 \chi -3)}{20 \chi}\ =\frac{(19 \lambda +16)}{20(\lambda +1)}\
    =\frac{4}{5} {d_T},,
\end{eqnarray}
and
\begin{eqnarray}
    \label{eq rheology D1  coeff clean}
    D_1(\lambda)=& (-3888 -27308 \chi +231041 \chi^2 -33637 \chi^3 \nonumber \\ &  -189761 \chi^4+159201 \chi^5)/(35280 \chi^4)\,,
\end{eqnarray}
and in Eq.~\refeq{eq shear N2 clean}, the second-order deformation analysis for clean drops includes
\begin{equation}
    \label{eq shear N2 clean coeff}
  D_{2}(\lambda)= \frac{3[12+9(1+\lambda)-25 (1+\lambda)^2]}{28 (1+\lambda)^2}
\end{equation}
Likewise in Eq.~\refeq{eq shear stress 12 surfactants} for droplets covered with insoluble surfactants \cite{vlahovska2009small},
\begin{equation}
    \label{eq rheology D  coeff surfactant}
    D_{3}=\frac{5}{1176 \beta^2}[245 \chi + 98 \beta (3+\chi)+ \beta^{2} ( -1059 + 1127\chi)]\,,
\end{equation}
where $\chi=1+\lambda$ and $\beta=CaMa$.

\subsection{Droplets with viscous interface} \label{appendix: viscous interface}
Coefficients needed in the analytical formulas for \\ droplets covered with viscous interfaces are listed in this appendix, for completeness. A full analysis for small deformation analytical results in shear and extensional flows are listed in Ref. \cite{narsimhan2019shape}. The coefficient in inclination angle formula Eq.~\refeq{eq inclination angle viscous Ca} in the limit when the small parameter $\epsilon=Ca$ is
\begin{eqnarray}
    \label{eq angle coeff viscous eps}
   a_D=& [-8(6 Bq_{d}+4 Bq_{s}+5\lambda +5)]/(64 Bq_{d}\nonumber \\ &+48 Bq_{s}  + 89 \lambda +46 Bq_{d} \lambda+52 Bq_{s}\lambda \nonumber \\ &+38 \lambda^2+32Bq_{d} Bq_{s} +48)\,.
\end{eqnarray}
The coefficients appearing in Eq.~\refeq{D Taylor viscous} in the limit when $\epsilon=\lambda^{-1}$ and $Bq_{s}\sim Bq_{s}=O(1)$,
\begin{eqnarray}
    \label{eq angle coeff viscous lam}
   \hat{a}_D&=-\frac{20}{19} \epsilon\,,
\end{eqnarray}
\begin{eqnarray}
    \label{eq angle coeff viscous lam 2 }
   \hat{a}_E&=\frac{5}{2}\epsilon - \epsilon^2 (\frac{15}{4} +\frac{5}{38}Bq_d +\frac{45}{19}q_s)\,,
\end{eqnarray}
and when $\lambda=O(1)$ and $\epsilon=Bq_{i}^{-1}$ where $i=s,d$ and $Bq_{s}\sim Bq_{d}$ are 
\begin{eqnarray}
    \label{eq angle coeff viscous lam 3}
   \hat{a}_D&=-\epsilon \left(\frac{3}{2} +\frac{Bq_s}{Bq_d}\right)\,,
\end{eqnarray}
\begin{eqnarray}
    \label{eq angle coeff viscous lam 4 }
   \hat{a}_E&=\frac{5}{4}\epsilon \left(3 +\frac{Bq_s}{Bq_d}\right)-\frac{5}{64} \epsilon^2 \left[96 +69 \lambda + (72+63 \lambda) \nonumber \right. \\ & \left. \times(Bq_s/Bq_d)+(24+26\lambda)(Bq_s/Bq_d)^2\right]\,.
\end{eqnarray}

\subsection{Droplets with interfacial slip} \label{appendix: slip interface}
The coefficients appearing in Eqs.~\refeq{eq viscosity slip}-\refeq{eq shear N1 N2 slip} for the viscometric functions of droplets with slip in shear flow are \cite{ramachandran2012effect}
\begin{equation}
    \label{eq coeff f slip droplet}
    f=\frac{1}{40}\left[\frac{\lambda(80 \bar \alpha+19)+16}{\lambda(5 \bar \alpha +1)+1}\right]^2\,,
\end{equation}
and 
\begin{eqnarray}
    \label{eq coeff g slip droplet}
   g &=\left(3\left[\lambda(80 \bar \alpha+19)+16\right]\left[5 \lambda^2(20\bar \alpha^2+4 \bar \alpha+5)+4 \right. \right. \nonumber \\  & \left. \left. +\lambda(40\bar \alpha+41)\right]\right)/\left(140\left[\lambda(5\bar \alpha +1)+1\right]^3\right)\,,
\end{eqnarray}
where $\bar \alpha$ is the dimensionless slip coefficient defined in section \ref{sec: relevant parameters}.

\section{Constitutive equation for dilute emulsion of clean droplets} \label{Appendix: Frankel Acrivos}
In this Appendix we present a discussion and list the main equations for a constitutive equation for a dilute emulsion of nearly spherical, clean drops with finite surface tension following the work by Frankel \& Acrivos \cite{frankel1970constitutive}. Under small deformation conditions, the particle extra stress contribution as shown in Eq.~\refeq{eq particle stress} reduces to 
\begin{eqnarray}
    \label{eq particle stress Frankel Acrivos}
     \Sigma^{p}_{ij} =& \mu_0 \phi \bigg\{ \frac{10 (\lambda - 1)}{2\lambda + 3} E_{ij} - \frac{24}{2\lambda + 3} F_{ij}
\nonumber \\ &  + \frac{360 (\lambda - 1)^2}{7(2\lambda + 3)^2} \tau_\sigma \mathcal{L}[F_{ip} E_{pj}] \nonumber \\ & 
+ \frac{288 (\lambda - 6)}{7(2\lambda + 3)^2} \tau_\sigma \mathcal{L}[F_{ip} F_{pj}] + O(\dot{\gamma}Ca^{2}) \bigg\}\,,
\end{eqnarray}
when $\epsilon=Ca$ and $\lambda$ is arbitrary, where $\tau_\sigma=\mu a/\sigma$ is the material relaxation time, $F_{ij}$ are the components of a second-order tensor that defines a shape distorting parameter f, such that, to leading order according to Eq.~\refeq{eq drop surface},
\begin{equation}
    \label{eq relation F and A}
    f=F_{ij} \left( \frac{\partial^2 r^{-1}}{\partial x_i \partial x_j}\right)_{r=a}= \frac{\bx\cdot \mathbf{A} \cdot \bx}{r^2} \,.
\end{equation}
The coefficient of the shape distorting parameter satisfies an evolution equation,
\begin{eqnarray}
    \label{eq Fij Frankel Acrivos}
    F_{ij} &+  \displaystyle g_1(\lambda)\tau_{\sigma}\frac{\mathcal{D}F_{ij}}{\mathcal{D}t}=g_2(\lambda) E_{ij}+g_3(\lambda)\tau_{\sigma} \mathcal{L}[F_{ip}E_{pj}] \nonumber \\ & +
    g_4(\lambda) \tau_{\sigma} \mathcal{L}[F_{ip}F_{pj}] + O(\dotgamma Ca^2)\,,
\end{eqnarray}
where 
\begin{equation}
    \label{eq symmetric traceless operator}
    \mathcal{L}M_{ij} = \frac{1}{2}\left(M_{ij}+M_{ji}\right) -\delta_{ij} M_{ll} \,,
\end{equation}
is a linear operator that yields the symmetric, traceless part of a second order tensor $\mathbf{M}$, $\delta_{ij}$ is the Kronecker delta, and the operator $\mathcal{D}/\mathcal{D}t$ is the co-rotational, Jaumann derivative that translates and rotates with the particle \cite{goddard1966inverse},
\begin{eqnarray}
    \label{eq Jaumann}
    \frac{\mathcal{D}M_{ij}}{\mathcal{D}t}=\frac{\partial M_{ij}}{\partial t} + u_k \frac{\partial M_{ij}}{\partial x_k}+W_{ik}M_{kj}-W_{kj}M_{ik}
\end{eqnarray}
where $\mathbf{W}$ is the vorticity tensor. The functions appearing in Eq.~\refeq{eq Fij Frankel Acrivos} are listed in the Appendix. In the limit when $\epsilon=\lambda^{-1}\ll 1$ for arbitrary, finite $Ca$, Eqs. \refeq{eq particle stress Frankel Acrivos} and \refeq{eq Fij Frankel Acrivos} reduced to 
\begin{eqnarray}
    \label{eq particle stress Frankel Acrivos lambda}
     \Sigma^{p}_{ij} &=\mu_0 \phi \bigg\{ 5 E_{ij} +12 (\tau_\sigma \lambda)^{-1} F_{ij} \nonumber \\ & + \frac{90}{7} \lambda^{-1} \mathcal{L}[F_{ip} E_{pj}] + O(\dot{\gamma}\lambda^{-2}) \bigg\}\,,
\end{eqnarray}
and 
\begin{eqnarray}
    \label{eq Fij Frankel Acrivos lambda}
    & \frac{20}{19} (\tau_\sigma \lambda)^{-1} F_{ij} +  \displaystyle \frac{\mathcal{D}F_{ij}}{\mathcal{D}t} =  
 \nonumber \\ & (5/6) E_{ij} -(10/7) \lambda^{-1}\mathcal{L}[F_{ip}E_{pj}] + O(\dotgamma \lambda^{-2})\,.
\end{eqnarray}

The constitutive equations \refeq{eq particle stress}, \refeq{eq particle stress Frankel Acrivos}, and \refeq{eq particle stress Frankel Acrivos lambda} are the main results of Frankel \& Acrivos small deformation theory for an emulsion of clean droplets. These results are in agreement with constitutive equations derived for suspensions of solid elastic \cite{goddard1967nonlinear} and for emulsions with an intrinsic material relaxation time \cite{oldroyd1958nonNewt}.

For steady or weakly time-dependent flows, in the limit when $\epsilon=Ca \ll 1$ and $\tau_\sigma \vert \mathbf{E} \cdot \mathbf{E}\vert \ll \vert \mathbf{E}\vert$, Eq.~\refeq{eq particle stress Frankel Acrivos} can be recast in the form of Oldroyd's constitutive equation \cite{oldroyd1958nonNewt}, 
\begin{eqnarray}
\label{eq constitutive Oldroyd form}
    & \displaystyle \Sigma_{ij} + \Lambda \frac{\mathcal{D}\Sigma_{ij}}{\mathcal{D}t} =
- p \delta_{ij}+ 2 \mu \eta_r \left(E_{ij} +\Lambda \frac{\mathcal{D}E_{ij}} {\mathcal{D}t}\right) \nonumber \\  & \displaystyle 
+\Lambda \mu \phi \left\{ - \frac{8}{5} \left(\frac{\Lambda}{\tau_\sigma}\right)\frac{\mathcal{D}E_{ij}} {\mathcal{D}t} \right. \\ & \left.  
\displaystyle + \frac{1}{(2\lambda+3)} \left[\frac{150}{7} g_T^2(\lambda)+18\frac{\lambda}{(\lambda+1)^2}\right] \mathcal{L}\left[ E_{ip} E_{pj}\right] \right\} \nonumber \,
\end{eqnarray}
where $\eta_r$ is the emulsion relative viscosity, $\mathcal{L}$ is a symmetric, traceless operator given by Eq. \refeq{eq symmetric traceless operator}, $\mathcal{D}/\mathcal{D}t$ is the Jaumman derivative defined in Eq.~\refeq{eq Jaumann}, $\Sigma_{ij}$ are the components of the volume-averaged stress tensor defined in Eq.~\refeq{eq particle stress}, and the timescale computed using
\begin{eqnarray}
    \label{eq Lambda relaxation time 1}
    \Lambda = \frac{(2\lambda + 3)(19\lambda + 16)}{40(\lambda + 1)} \tau_\sigma = c_1(\lambda)^{-1}\tau_\sigma \,,
\end{eqnarray}
represents a material relaxation time due to finite values of surface tension which recovers the shape relaxation time defined in  Eq.~\refeq{eq evolution distortion} in terms of the coefficient $c_1(\lambda)$ multiplying the distortion tensor $\mathbf{A}$. Following Oldroyd's findings \cite{oldroyd1958nonNewt}, Eq.~\refeq{eq constitutive Oldroyd form} describes a viscoelastic fluid with non-zero normal stress differences, a shear-rate dependent relative viscosity, and a positive Weissenberg effect \cite{frankel1970constitutive}. These results are confirmed by second-order small deformation theory given by Eqs.~\refeq{eq shear stress 12 clean}, \refeq{eq shear N1 clean}, and \refeq{eq shear N2 clean}.

Fundamental aspects of rheological constitutive equations of state, such as invariance properties, have been discussed in a classical work by Oldroyd \cite{oldroyd1950formulation} for dilute emulsions. At sufficiently weak flows with small enough velocity gradients, the last term in Eq.~\refeq{eq constitutive Oldroyd form} has a higher-order contribution to the total particle stress and Eq.~\refeq{eq constitutive Oldroyd form} reduces to a three-parameter, Jeffrey-like constitutive equation given by Eq.~\refeq{eq constitutive Jeffreys form} in the main text.


\section{Data used in Figs.~\ref{FIG:3}, ~\ref{FIG:Effective viscosity comparison} and \ref{FIG:6}}  
\label{Appendix: dataset fig 3}

\subsection{Experimental datasets used in Fig.~\ref{FIG:3}} 

The schematics redrawn and used in Fig.~\ref{FIG:3} are adapted from the experimental results detailed in Ref.~\cite{rumscheidt1961particle}.

\begin{itemize}
    \item First row: $\lambda = 6$. Silicone oil 30,000 (Dow Corning fluid) in 60 cP oxidized castor oil (Pale 4). Interfacial tension 6.0 dyn/cm.

    \item Second row: $\lambda = 1$. Oxidized castor oil (Pale 4) in 52.6 cP silicone oil 5000 (Dow Corning fluid). Interfacial tension 4.8 dyn/cm.

    \item Third row: $\lambda = 0.7$. Oxidized castor oil (Pale 4) in 90 cP corn syrup. Interfacial tension 21 dyn/cm.

    \item Fourth row: $\lambda = 0.0002$. Distilled water in 52.6 cP silicone oil 5000 (Dow Corning fluid). Interfacial tension 38 dyn/cm.
    
\end{itemize}

\subsection{Figure~\ref{FIG:Effective viscosity comparison}: $\eta_r$ vs $\phi$}
\textbf{Datasets:}
\begin{itemize}
    \item Squares: obtained from Fig.~8 in Ref.~\cite{mason1996yielding} for a monodisperse silicon oil-in-water emulsion with SDS concentration of $10 $mM, droplet size $a = 0.55 \mu$m, viscosity of the oil $\lambda \mu =12$ cP, water viscosity $\mu=0.997$ cP, $\lambda=12$, and $\sigma=9.8$ dyn/cm.
    
    \item Circles: obtained from Fig.~8 in Ref.~\cite{mason1996yielding} for a monodisperse silicon oil-in-water emulsion with SDS concentration of $10$ mM, droplet size $a = 0.20 \mu$m, viscosity of the oil $\lambda \mu =12$ cP, water viscosity $\mu=104$ cP, $\lambda=0.12$, and $\sigma=9.8$ dyn/cm.
    
    \item Pentagons: obtained from Fig.~6 set 2 in Ref.~\cite{pal2000shear} for a polydisperse petroleum oil-in-water emulsion with Triton-X-100 concetration of 2.1 wt\%. Effective drop radius $a_{32}=9.12\,\mu$m, viscosity of the oil $\lambda \mu=5.52$ cP, water viscosity $\mu=0.997$ cP, $\lambda=5.54$, and $\sigma=1.5$ dyn/cm.
    
    \item Triangles: obtained from Fig.~6 set 2 in Ref.~\cite{pal2000shear} for a polydisperse petroleum oil-in-water emulsion with Triton-X-100 concetration of 2.1 wt\%. Effective drop radius $a_{32}=9.12\,\mu$m, viscosity of the oil $\lambda \mu=5.52$ cP, water viscosity $\mu=0.997$ cP, $\lambda=5.54$, and $\sigma=1.5$ dyn/cm.
\end{itemize}

\textbf{Models:} 
\begin{itemize}
    \item Taylor \cite{taylor1932viscosity}: using Ref.~\cite{pal2000shear} emulsion of oil and water viscosities $\lambda \mu=5.52$ cP and $\mu = 0.997$ cP, respectively, and $\lambda=5.54$.
    
    \item Choi \& Schowalter \cite{choi1975rheological}: using Ref.~\cite{pal2000shear} emulsion of oil and water viscosities $\lambda \mu =5.52$ cP and $\mu = 0.997$ cP, respectively, and $\lambda=5.54$.

    \item Krieger-Dougherty-like: using $\phi_m=0.64$ and $\lambda=5.54$ according to Ref.~\cite{pal2001novel}. Another line was plotted with $\lambda=110$ from Ref.~\cite{mason1996yielding} to better overlap the experimental data.
\end{itemize}

\subsection{Figure \ref{FIG:6}: $\bar G$ vs $\phi$}
\textbf{Datasets:}
\begin{itemize}
    \item Triangles: $E(\phi)$ points extracted from Fig.~6 in Ref.~\cite{princen1986rheology}, where $E(\phi)=G \sigma/(a_{32} \phi^{1/3})$. Polydisperse paraffin oil-in-water emulsion with 11.6 wt\% Alipal CD-128, 58\% active. Each emulsion has an individual mean diameter and interfacial tension as follows: $a_{32} = 8.43-8.92\,\mu$m, $\lambda \mu = 49.2$ cP,  $\mu = 2.22$ cP, $\lambda=22.2$, and $\sigma=6.20-6.86$ dyn/cm.

    \item Circles: dataset for particle size $a=0.53\mu$m from Ref.~\cite{lacasse1996model} according to Fig.~2b (black down-triangles) in Ref.~\cite{KimMason017advances}. Monodisperse silicon oil-in-water emulsion with SDS concentration of C = 10 mM, where $\lambda \mu = 110$ cP, $\mu= 0.997$ cP, $\lambda=110$, and $\sigma = 9.8$ dyn/cm.
\end{itemize}

\textbf{Models:}
\begin{itemize}
    \item Princen \& Kiss \cite{princen1989rheologyIV} model plotted in the range $\phi=0.73-1$.
    \item Wilking \& Mason \cite{wilking2007irreversible} model plotted in the range $\phi=0.73-1$; assuming $\phi_{m}=0.71$.
\end{itemize}

\subsection{Figure~\ref{FIG:6}: $\bar \tau_Y$ vs $\phi$}
\begin{itemize}
    \item Squares: rescaled data from Table 1 of Ref.~\cite{princen1989rheologyIV}. Polydisperse paraffin oil-in-water emulsion with 10\% Neodol 25-3S + 2\% Neodol 25-9. Each emulsion has an individual mean diameter and interfacial tension in the ranges: $a_{32}=5.73-10.2 \mu$m, oil viscosity $\lambda \mu=49.2$ cP, water viscosity $\mu=1.53$ cP, $\lambda=32.2$, and $\sigma=4.50-4.92$ dyn/cm.
    
    \item Triangles: replotted from Fig.~4 (circles) in Ref.~\cite{mason1996yielding}. Monodisperse silicon oil-in-water emulsion with SDS concentration of $10$ mM, drop size $a=0.25\mu$m, $\lambda \mu = 12$ cP, $\mu=104$ cP, $\lambda=0.12$, and interfacial tension $\sigma = 9.8$ dyn/cm.

    \item Hexagons: replotted from Fig.~4 (squares) in Ref.~\cite{mason1996yielding}. Monodisperse silicon oil-in-water emulsion with SDS concentration of $10$ mM, drop size $a=0.53 \mu$m, $\lambda \mu = 12$ cP, $\mu=104$ cP, $\lambda=0.12$, and interfacial tension $\sigma = 9.8$ dyn/cm.
\end{itemize}

\textbf{Models:}
\begin{itemize}
    \item Princen \& Kiss (1989) \cite{princen1989rheologyIV}: plotted in the range $\phi=0.646-1$. Scaled with the following parameters:
	$a_{32} = 10.05\mu$m, $\sigma = 4.723$ dyn/cm.

 \item Mason, Bibette and Weitz (1996) \cite{mason1996yielding}: plotted in the range $\phi=0.646-1$ using the empirical quadratic fit for the scaled yield stress $\tau_Y/(\sigma/a)=0.51 (\phi_{eff}-\phi_c)^2 $ where $\phi_c=0.62$.

\end{itemize}

\section{Table of symbols and nomenclature} \label{appendx: symbols nomenclature}
Table \ref{FIG:Table of symbols} is like a Rosetta stone that lists the symbols used in this work against the nomenclature adopted by the Society of Rheology \cite{SOR2013official}. We have endeavored to adopt consistent symbols within the paper and introduced simplifications when possible for brevity and improving clarity.   


\begin{table}[h!]
\centering
\setlength{\arrayrulewidth}{0.5mm} 
\setlength{\tabcolsep}{5pt} 
{\rowcolors{3}{green!80!yellow!50}{green!70!yellow!40}
\begin{tabular}{ |p{4.4cm}|p{1.25cm}|p{1.25cm}|  }
\hline
\multicolumn{3}{|c|}{List of Symbols} \\
\hline
Name & Our \,Symbol & SoR \,Symbol \\
\hline
Shear stress & $\mathbf{\tau}$ & $\mathbf{\sigma}$ \\
Yield stress & $\tau_Y$   & $\sigma_Y$ \\
Fluid viscosity & $\mu$ & $\eta_f$ \\
Maximum packing fraction & $\phi_m$ & $\phi_{max}$ \\
Local stress tensor & $\mathbf{T}$ & $\mathbf{\sigma}(\bx,t)$ \\
Boussinesq number & $B_q$ & $Bo$   \\
Interfacial shear viscosity & $\mu_s$ & $\mu^s$ \\
Interfacial dilatational viscosity & $\mu_d$ & $\kappa^s$ \\
Rate-of-strain tensor & $\mathbf{E}$ & $\mathbf{\dot{\gamma}}/2$ or $\mathbf{D}$ \\
\hline
\end{tabular}}
 \caption{Correspondence between symbols and nomenclature used in the review and official list of symbols and nomenclature of The Society of Rheology \cite{SOR2013official}.}
 \label{FIG:Table of symbols}
\end{table}

\section*{Acknowledgement}

For discussions on the key aspects of fundamental and applied emulsion rheology and feedback on early draft versions, the authors wish to acknowledge Stefan Baier, Hy Bui, Philip Erni, Peter Fischer, Marc-Antoine Fardin, Reza Foudazi, Shamsheer Mahammad, Taygoara Oliveira, Vivek Narsimhan, Joe Peterson, Arun Ramachandran, Naveen Reddy, Abhinendra Singh, Samanvaya Srivastava and Siva Vanapalli. We wish to dedicate the review to Andy Acrivos for Vivek's interactions with Andy during his sabbatical motivated him to dive through many fundamental papers included here. Rodrigo acknowledges the Department of Chemical Engineering and the Office of Diversity, Equity \& Engagement at UIC for funding and research support as a Bridge to Faculty Postdoctoral Fellow. Nadia acknowledges the Department of Chemical Engineering at UIC for teaching assistantship and The Kraft Heinz Company for funding.  


\clearpage
\bibliographystyle{elsarticle-num-names}

\bibliography{cas-refs}


\end{document}